\documentclass[iop]{emulateapj}
\usepackage{amsmath,amssymb,float}
\usepackage{rotating,listings,xfrac,url}
\renewcommand{\thefootnote}{$\dagger$} 

\slugcomment{Accepted by The Astronomical Journal 2016}
\shorttitle{Determining the Age of NGC 6819 with a New Triple Eclipsing System}
\shortauthors{Brewer et al.}

\begin{document}
 
\title{Determining the Age of the {\it{Kepler}} Open Cluster NGC 6819
With a New Triple System and Other Eclipsing Binary Stars
\footnotemark[$\dagger$]}\footnotetext[$\dagger$]{This is paper 57 of the WIYN Open Cluster Study (WOCS).}

\renewcommand{\thefootnote}{\arabic{footnote}}

\author{Lauren N. Brewer\altaffilmark{1}; Eric L. Sandquist\altaffilmark{1};
  Robert D. Mathieu\altaffilmark{2,9}; Katelyn Milliman\altaffilmark{2,9}; Aaron
  M. Geller\altaffilmark{3,4,9}; Mark W. Jeffries, Jr.\altaffilmark{1}; Jerome
  A. Orosz\altaffilmark{1}; Karsten Brogaard\altaffilmark{5,6}, Imants
  Platais\altaffilmark{7}; Hans Bruntt\altaffilmark{5}; Frank
  Grundahl\altaffilmark{5}; Dennis Stello\altaffilmark{8}; S$\o$ren
  Frandsen\altaffilmark{5}}

\altaffiltext{1}{\footnotesize{San Diego State University, Department of Astronomy, San Diego, CA 92182; ezereve@gmail.com; esandquist@mail.sdsu.edu; jorosz@mail.sdsu.edu; mjeffries11@gmail.com}}
\altaffiltext{2}{\footnotesize{University of Wisconsin-Madison, Department of Astronomy, Madison, WI 53706; mathieu@astro.wisc.edu; milliman@astro.wisc.edu}}
\altaffiltext{3}{\footnotesize{Center for Interdisciplinary Exploration and Research in Astrophysics (CIERA) and Northwestern University, Department of Physics and Astronomy, 2145 Sheridan Road, Evanston, IL 60208; a-geller@northwestern.edu}}
\altaffiltext{4}{\footnotesize{Department of Astronomy and Astrophysics, University of Chicago, 5640 S. Ellis Avenue, Chicago, IL 60637, USA}}
\altaffiltext{5}{\footnotesize{Stellar Astrophysics Centre, Department of Physics and Astronomy, Aarhus University, Ny Munkegade 120, DK-8000 Aarhus C, Denmark; kfb@phys.au.dk, bruntt@gmail.com, fgj@phys.au.dk, srf@phys.au.dk}}
\altaffiltext{6}{\footnotesize{Department of Physics \& Astronomy, University of Victoria, P.O. Box 3055, Victoria, BC V8W 3P6, Canada; }}
\altaffiltext{7}{\footnotesize{Department of Physics and Astronomy, The Johns Hopkins University, Baltimore, MD 21218; imants@pha.jhu.edu}}
\altaffiltext{8}{\footnotesize{Sydney Institute for Astronomy (SIfA), School of Physics, University of Sydney, NSW, 2006, Australia; stello@physics.usyd.edu.au}}
\altaffiltext{9}{\footnotesize{Visiting Astronomer, Kitt Peak National Observatory, National Optical
Astronomy Observatories, which is operated by the Association of Universities
for Research in Astronomy, Inc. (AURA) under cooperative agreement with
the National Science Foundation.}}

\begin{abstract}
As part of our study of the old ($\sim$2.5 Gyr) open cluster NGC 6819
in the {\it Kepler} field, we present photometric ({\it Kepler} and
ground-based $BVR_CI_C$) and spectroscopic observations of the
detached eclipsing binary WOCS 24009 (Auner 665; KIC 5023948) with a
short orbital period of 3.6 days. WOCS 24009 is a triple-lined system,
and we verify that the brightest star is physically orbiting the
eclipsing binary using radial velocities and eclipse timing variations.
The eclipsing binary components have masses $M_{B} =1.090 \pm 0.010
M_{\sun}$ and $M_{C} =1.075 \pm 0.013 M_{\sun}$, and radii $R_{B}
=1.095 \pm 0.007 R_{\sun}$ and $R_{C} =1.057 \pm 0.008 R_{\sun}$.  The
bright non-eclipsing star resides at the cluster turnoff, and ultimately its mass
will directly constrain the turnoff mass: our preliminary
determination is $M_{A} =1.251 \pm 0.057 M_{\sun}$.  A careful
examination of the light curves indicates that the fainter star in the
eclipsing binary undergoes a very brief period of total eclipse, which
enables us to precisely decompose the light of the three stars and
place them in the color-magnitude diagram.

We also present improved analysis of two previously discussed detached
eclipsing stars in NGC 6819 (WOCS 40007 and WOCS 23009) en route to a
combined determination of the cluster's distance modulus $(m-M)_V=12.38\pm0.04$. 
Because this paper significantly increases the number of measured stars in the cluster, we can better constrain
the age of the color-magnitude diagram to be $2.21\pm0.10\pm0.20$ Gyr.
Additionally, using all measured eclipsing binary star masses and radii, we constrain the age to
$2.38\pm0.05\pm0.22$ Gyr.
The quoted uncertainties are estimates of
measurement and systematic uncertainties (due to model physics
differences and metal content), respectively.
\end{abstract}


\section{Introduction}\label{intro}

Measurements of the masses and radii of the component stars in
detached eclipsing binaries (DEB) can be used to precisely determine
the age of the stars if at least one of the eclipsing stars has begun
to evolve away from the main sequence. The use of mass and
radius ($M-R$) measurements of eclipsing stars avoids or minimizes
systematic uncertainties introduced by factors such as distance,
interstellar reddening, and color-temperature conversions that can
affect age measurements \citep{andersen91,torres10}. When eclipsing
binaries occur in star clusters, their utility increases dramatically
because they place constraints on the age of {\it all} the stars
in the cluster. 
Multiple DEBs in a cluster can provide $M-R$ measurements for stars having a
range in mass, and can constrain the age even more tightly. With a
well-determined age for a cluster, the color-magnitude diagram (CMD)
can be used to test the effects of evolution on an even larger number
of stars.  Comparisons of the masses and radii of eclipsing stars and the
CMD with theoretical stellar models can also lead to deeper inferences
about chemical composition variables such as helium abundance
\citep{brogaard11,brogaard12}.

This study focuses on the rich open cluster NGC 6819, which has been
observed thoroughly from the ground and with the {\it Kepler} spacecraft.
NGC 6819 is a well-studied cluster: its CMD
\citep{burkhead71,lindoff72,auner74,rosvick98,kalirai01,yang} as well as stellar
variability \citep{kaluzny88,street02,street03,street05,talam10} have been
examined in multiple studies. In addition, the cluster has been part of the
WIYN Open Cluster Survey (WOCS; \citealt{mathieu00}) with
comprehensive long-term radial-velocity studies presented by \cite{hole09} and \cite{milliman}.  

Thanks to the nearly continuous monitoring and high precision
photometry by {\it Kepler}, studies of stellar ages using multiple
techniques have been kickstarted.  Because several techniques can be
applied within the same cluster, a comparison of results should
ultimately improve our understanding of each.  \cite{basu11} used
asteroseismic data from {\it Kepler} to constrain basic properties of
red giant stars and to place constraints on the cluster distance
[$(m-M)_{0} = 11.85 \pm 0.05$] and age (between 2.1 and 2.5 Gyr).
Stellar rotation measurements have been used to examine the
gyrochronological ages of two of the four open clusters in the {\it
  Kepler} field: NGC 6811 \citep{meibom11} and NGC 6819
\citep{meibom15}.  \cite{kalirai01} describes how the age of NGC 6819
can be found using the faint end of the white dwarf cooling sequence,
and \citet{bedin} present an age of $2.25\pm0.20$ Gyr using that
technique. For our part, we have so far analyzed two eclipsing systems
in NGC 6819 (WOCS 23009 / KIC 5024447, \citealt{sandquist13}; WOCS
40007 / KIC 5113053, \citealt{jeffries13}).  

In this paper we present the triple-lined system WOCS 24009 (also
known as KIC 5023948 / Auner 665;
$\alpha_{2000}=19^{\mathbf{h}}40^{\mathbf{m}} 57\farcs82$,
$\delta_{2000}=+40^{\circ}09\arcmin27\farcs4$) to add to the grid of cluster stars with
precisely-determined masses and to further constrain the age of NGC
6819. We identify the three stars as follows: component A is the
brightest contributor to spectra and is on a wide orbit around the
short-period detached eclipsing binary made up of components B and C,
where B is slightly brighter and more massive than C and is the star
being eclipsed during primary eclipse, as we will discuss in the
analysis sections.  This system is promising because all three stars
are bright enough to be studied spectroscopically, the three can be
shown to be physically associated, and it contains one of the most
evolved stars in the cluster whose mass can be directly determined
from radial-velocity measurements.

In \S 2, we describe the observations we employed, and describe the
improved spectroscopic analysis techniques that allowed us to minimize
the mass and radius measurement uncertainties in the the face of three
sets of blended lines.  We also used these techniques to improve our
previously presented measurements of WOCS 23009 and WOCS 40007. In \S
3, we discuss the membership of the system, and describe the modeling
of the eclipsing binary and the orbit of the bright non-eclipsing
star. In \S 4, we discuss the photometry of the stars, and use all
three eclipsing systems in a combined analysis to derive the cluster distance modulus and age.

\setcounter{footnote}{9}

\section{Observations and Data Reduction}

Our datasets include ground-based and space-based {\it Kepler} photometric
observations along with long-term radial-velocity observations. In addition to
the work on the WOCS 24009 system, we also discuss improvements to the dataset
for WOCS 40007 \citep{jeffries13} below.

\begin{center}
\begin{deluxetable}{lcccc}[!h]
\tablewidth{0pt}
\tablecaption{Observed Nights at the Mount Laguna Observatory 1-meter telescope \label{dates}}
\tablehead{
\colhead{Date} & \colhead{Filter} & \colhead{mJD Start\tablenotemark{a}} & 
\colhead{$t_{exp}$ (s)} & \colhead{$N$}}
\startdata
2008 June 15 & $B$ & 4633.699 & 180 & 22 \\
 2009 June 12 & $V$ & 4994.701 & 120 & 147\\
 2009 Oct. 22 & $V$ & 5126.589 & 120 & 59\\
2010 June 22 & $B$ & 5370.708 & 600 & 34\\
2010 Sep. 14 & $B, R_c$ & 5454.781 & 600 & 7, 7\\
2011 May 22 & $B, V$ & 5704.802 & 600, 600 & 6, 9\\
2011 June 02 & $B, V$ & 5715.738 & 600, 600 & 9, 12\\
2011 June 11 & $V$ & 5724.725 & 600 & 35\\
2011 July 03\tablenotemark{d} & $I_c$ & 5746.677 & 300 & 41\\
2011 July 14 & $R_c$ & 5757.660 & 420 & 62\\
2011 July 23\tablenotemark{m} & $I_c$ & 5766.657 & 300 & 83\\
2011 July 25\tablenotemark{d,m} & $B, I_c$ & 5768.662 & 300, 300 & 33, 41\\
2011 Aug. 14 & $R_c$ & 5788.639 & 420 & 65\\
2011 Aug. 25 & $B, V$ & 5799.627 & 600, 600 & 13, 21\\
2011 Sep. 16 & $B, V$ & 5821.631 & 600, 600 & 17, 15\\
2011 Oct. 17 & $B, V$ & 5852.582 & 300, 300 & 28, 20\\
2011 Oct. 28 & $I_c$ & 5863.587 & 300 & 44\\
2011 Nov. 28 & $V$ & 5894.557 & 300 & 34\\
2012 Apr. 18 & $V, I_c$ & 6036.838 & 300 & 18, 20\\
2014 June 8 & $B, V$ & 6817.709 & 300, 300 & 45, 16 \\
2014 July 20 & $B$ & 6859.669 & 600 & 42 
\enddata
\tablenotetext{a}{mJD = BJD - 2 450 000}
\tablenotetext{d}{Data reduction used dome flat field images.}
\tablenotetext{m}{Data reduction used morning twilight flat field images.}
\end{deluxetable} 
\end{center}

\subsection{Ground-based Photometry}\label{phot}

We obtained images in $B, V, R_{c}$, and $I_{c}$ filters using the
Mount Laguna Observatory (MLO) 1 m telescope.  WOCS 24009 eclipse light curves
were computed from two nights in 2010 in addition to dedicated nights
in 2011, 2012, and 2014.  The observations on nights of eclipses are
given in Table \ref{dates}. The dataset also includes earlier images
presented by \cite{talam10}. Exposure times were typically 300 or 600
seconds for $B$ and $V$, 420 seconds for $R_{c}$, and 300 seconds for
$I_{c}$. The CCD 2005 camera covers $13\farcm7\times13\farcm7$ on sky
with a scale of $0\farcs41$ per pixel.

All ground-based observations were overscan corrected, bias subtracted
using a master bias computed from twenty individual images, and
calibrated using nightly master flat-field frames to correct
pixel-to-pixel sensitivity variations.  Evening twilight flats were
generally used, but nights for which we utilized dome flats or morning
flats are noted in Table \ref{dates}. The image processing tasks were
completed using basic procedures in IRAF\footnote{IRAF is
  distributed by the National Optical Astronomy Observatory, which is
  operated by the association of Universities for Research in
  Astronomy, Inc., under cooperative agreement with the National
  Science Foundation.}.  We note that since the \citet{jeffries13}
paper it was discovered (D. Leonard, private communication) that a
nonlinearity in the CCD response was present as a result of improperly
set voltages on readout transistors that were used between December
2008 and November 2013. The following correction
\begin{align}
& ADU_{cor} = ADU * (1.01353 - 0.115576 \cdot (ADU / 32767) \nonumber \\
& + 0.0296378 \cdot (ADU / 32767)^2) \nonumber
\end{align}
was applied to pixel counts on re-reduced flat field and object images
(after overscan and bias subtraction) taken during that time frame
using the IRAF task {\it irlincor}. The nonlinearity had small but
noticeable effects on photometric scatter and trends within nights of
data but minor effects on measured stellar characteristics.

Photometric measurements were made using the 
\begin{figure*}[!ht]
\begin{center}
\includegraphics [width=0.9\textwidth]{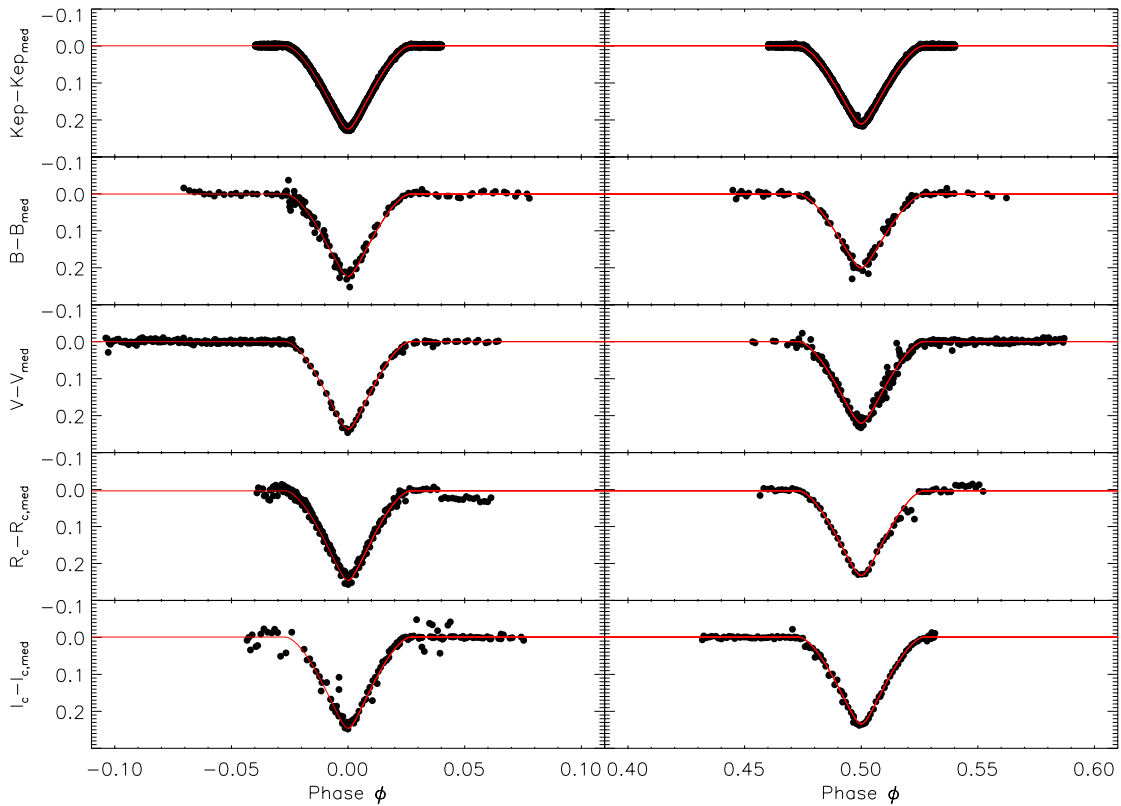}
\vspace{-20pt}
\caption{Photometry for WOCS 24009 from the {\it Kepler} spacecraft, and ground-based $B$,
  $V$, $R_{c}$, and $I_{c}$ observations, compared with the best-fit ELC model
  solution ({\it solid line}) for phases in and immediately surrounding the
  eclipses.
  \label{lc}}
\end{center}
\end{figure*}DAOPHOT II/ALLSTAR suite
of programs \citep{daoph}. Because crowding was an issue for a subset
of the stars, we conducted point-spread function (PSF) photometry on
all of the ground-based images. Typically 100-150 bright stars were
chosen to determine the PSF and its variation across each frame. To
improve the quality of the photometry in a differential sense, we used
the observed ensemble of stars to iterate to a solution for median
magnitudes of all stars and photometric zeropoints for each image
\citep{s1082,honey}.  Figure \ref{lc} shows WOCS 24009 primary and secondary
eclipse observations phased to the binary period with the slightly
deeper eclipse centered on phase $\phi = 0$.

\subsection{{\it Kepler} Photometry} 

The {\it Kepler} spacecraft made nearly continuous photometric
observations of a field between Cygnus and Lyra that included NGC 6819
--- mission details have been previously presented in
\cite{borucki10}, \cite{koch10}, \cite{batalha10}, \cite{caldwell10},
and \cite{gilliland10}.  ``Superaperture'' observations of the cluster
were made as part of the Kepler Cluster Study \citep{meibom11}. These
involved a series of 20 image stamps covering the center of NGC 6819,
together forming a 200 $\times$ 200 pix square field of view with
$13\farcm25$ on a side. Exposures for these stamps were taken in the
long-cadence mode of 30 minutes.  Eclipses of the binary components
were observed using {\it Kepler} data for quarters 1-16, excluding
quarters 6, 10, and 14 when the cluster fell on an unresponsive CCD
module.  We retrieved the public data from the Mikulski Archive for
Space Telescopes (MAST), operated by the Space Telescope Science
Institute.\footnote{\tt
  http://archive.stsci.edu/kepler/kic10/search.php} Data processing
was done using a series of Python tasks based on PyKe\footnote{\tt
  http://keplergo.arc.nasa.gov/PyKE.shtml} software tools to reduce
and analyze the {\it Kepler} simple aperture photometry (SAP). We
derived times of mid-eclipse for all observable eclipses, making use
of one month of short-cadence data in quarter 12 for its higher time
resolution. We restricted our later analysis to photometry in and
immediately around times of eclipse.

For {\it Kepler} photometry it is necessary to ensure that
instrumental trends are removed and that contamination from other
stars are accounted for.  WOCS 24009 is located just outside of the
center of the cluster (about $4\farcm4$ away) but is very close to neighboring stars. With a
large pixel size of 4$\arcsec$ across, contamination from other stars
seemed likely. Contamination values as determined by the {\it Kepler}
team ranged from 9.1\% to 15.3\% during different observing
seasons. Even after correcting for different contamination values,
cotrending the photometry against the light curves of other stars on
the CCD, and fitting and removing effects from star spots, we still
found significant variations in observed eclipse depths during a
single quarter of observations and from quarter-to-quarter.
In the end, we opted to use the short-cadence data from quarter
12 for our analysis due to the higher time resolution and in order to avoid
issues of inconsistent eclipse depth from quarter to quarter. We did not apply
any crowding corrections to this dataset as this will be fit as a free
parameter in the binary star models (see \S \ref{ELC}). 
The top panel in Figure \ref{lc} shows the trimmed, phased light curves of the short-cadence
Kepler data.

\begin{deluxetable}{cccl}[!htp]
\tablewidth{0pt}
\tablecaption{Eclipse Timings \label{etimes}}
\tablehead{\colhead{BJD - 2450000} & \colhead{$\sigma$} & 
\colhead{Eclipse} & \colhead{Notes}}
\startdata
\multicolumn{4}{c}{WOCS 24009}\\
2104.86904 & 0.00027 & P & $R_{c}$ ground\\
4697.66460 & 0.00033 & S & $R_{Kron}$ ground\\
4965.88626 & 0.00025 & P & Beginning of Kepler Q1\\
4967.71076 & 0.00019 & S & \\
4969.53563 & 0.00022 & P & \\
4971.36016 & 0.00017 & S & \\
4973.18516 & 0.00017 & P & \\
4975.00933 & 0.00025 & S & \\
4976.83423 & 0.00008 & P & \\
4978.65877 & 0.00024 & S & \\
4980.48360 & 0.00020 & P & \\
\multicolumn{4}{c}{WOCS 40007}\\
2102.88901 & 0.00008 & P & $R_{c}$ ground\\
4623.88162 & 0.00015 & S & V  ground\\
4698.73250 & 0.00011 & P & $R_{Kron}$ ground\\
4964.69147 & 0.00025 & S & Beginning of Kepler Q1\\
4966.28385 & 0.00023 & P &\\
4967.87663 & 0.00019 & S &\\
4969.46896 & 0.00013 & P &\\
4971.06181 & 0.00005 & S &\\
4972.65416 & 0.00012 & P &\\
4974.24685 & 0.00022 & S &\\
4975.83930 & 0.00022 & P &
\enddata
\tablecomments{This table is published in its entirety in the electronic edition, but a portion is shown here for guidance regarding its form and content.}
\end{deluxetable}

For the triple system WOCS 40007, we followed similar procedures, deriving
eclipse times (see Table \ref{etimes}) to measure orbit parameters for the
fainter non-eclipsing star, and using short-cadence data from quarter 12 in an update of
our earlier modeling of ground-based light curves \citep{jeffries13}.

\subsection{Spectroscopy}\label{spectra}

Radial velocities (RVs) were obtained using spectra from the high-precision survey of NGC
6819 \citep{hole09,milliman}.  These spectroscopic data were taken as part of the WIYN
Open Cluster Survey (WOCS; \citealt{mathieu00}) using the WIYN\footnote{The 
WIYN Observatory is a joint facility of the University of Wisconsin-Madison, 
Indiana University, the National Optical Astronomy Observatory and the 
University of Missouri}
3.5 m telescope on Kitt Peak and Hydra multi-object
spectrograph instrument (MOS), which is a fiber-fed spectrograph capable of
obtaining about 80 spectra simultaneously ($\sim$10 fibers set aside for sky
measurements and $\sim$70 for stellar spectra). Observations used the echelle
grating with a spectral resolution $\sim$ 15 km s$^{-1}$. Most of the spectra
were centered at 513 nm with a 25 nm range covering an array of narrow
absorption lines around the Mg b triplet. Spectroscopic observations for
WOCS 24009 were generally completed using 2 hour integrations per visit that
were split into three 2400 s integrations to allow for rejection of cosmic
rays.

Image processing was done using standard spectroscopic {\it IRAF}
procedures. First, science images were bias and sky subtracted, and
then the extracted spectra were flat fielded, throughput corrected,
and dispersion corrected.  The spectra were calibrated using one 100 s
flat field and two bracketing 300 s ThAr emission lamp spectra for
each observation.  Once the RVs were derived (see below), we also
corrected for individual fiber offsets present in the Hydra MOS. A
more detailed explanation of the acquisition and reduction of WOCS
observations can be found in \cite{geller08}.

\subsubsection{Spectral Disentangling and Broadening Function Analysis}\label{bf}

The spectra of WOCS 24009 are fairly complex to analyze given that there are
three cluster member stars contributing absorption-line features.  The
brightest component does not belong to the eclipsing binary and there is
substantial blending of the lines when the stars are near their eclipse times.
The need to derive precise RVs in such situations led us to use
broadening functions (BF), which have been demonstrated to produce reliable
measurements even for contact binaries with a great deal of rotational
broadening \citep{bf92,bf02}. BFs also minimize systematic effects on the
measurements like ``peak pulling'' in situations where traditional techniques
such as cross correlation are unable to cleanly separate the contributions from
different components \citep{bf02}.

Briefly, the broadening function can transform a sharp-lined synthetic or star
spectrum into a broadened spectrum through a convolution. The broadening
function contains information on how the intensity of the star's light is
redistributed due to Doppler effects resulting from galactic motion, orbital
motion, or rotation, among others. As a result, significant peaks in the
broadening function give a measure of the overall RVs of the
components, the function shape is an indicator of the rotation, and the total
area of the peak is a proxy for luminosity in the spectral region observed.

However, even with this improvement, line blending prevented optimal
determination of RVs for some epochs. As a result, we
also applied the technique of spectral disentangling
\citep{gonzo}. Because the spectra of the individual stars can be
derived when there are a sufficient number of spectra taken with good
(and differing) radial-velocity separation between the components, it
is sometimes possible to derive RVs even with fairly
severe line blending.  Starting with initial guesses for the radial
velocities, average spectra for the individual stars can be derived
and used to subtract out the lines of all but one of the component
stars in turn, and the subtracted spectra can then be used to make
cleaner measurements of the RVs. The average stellar
spectra and RVs can then be iteratively improved, and
convergence tends to be rapid.  In our case, there are three
detectable stars in the spectra, and we extended the \citet{gonzo}
algorithm to iteratively disentangle all three, starting with radial
velocity estimates from broadening function analysis of the entangled
spectra.

\begin{deluxetable}{lrrr}[!htp]
\tablewidth{0pt}
\tablecaption{Measured Radial Velocities from the WIYN 3.5 m Telescope\label{rvtab}}
\tablehead{\colhead{mJD\tablenotemark{a}} & \colhead{$v_A$ (km s$^{-1}$)} & \colhead{$v_B$ (km s$^{-1}$)} & \colhead{$v_C$ (km s$^{-1}$)}
}
\startdata
\multicolumn{4}{c}{WOCS 24009}\\
\hline
2461.8591 & $4.14 \pm 0.06$ & $83.60 \pm 0.36$ & $-81.77 \pm 0.54$\\
2542.6483 & $5.89 \pm 0.25$ & & $-23.05 \pm 4.54$\\
2840.9238 & $8.51 \pm 0.37$ & $77.28 \pm 3.21$ & $-91.44\pm 4.25$\\
2857.7341 & $6.98 \pm 0.26$ & $-89.59 \pm 1.34$ & $83.54\pm 1.78$\\
3190.8469 & $12.22 \pm 0.81$ & $46.72 \pm 3.22$ & $-51.54 \pm 2.54$\\
3576.8340 & $12.65 \pm 0.19$ & $-71.66 \pm 1.64$ & $62.93 \pm 1.75$\\
3810.9662 & $12.06 \pm 0.53$ & & $-10.26\pm 4.77$\\
\hline
\multicolumn{4}{c}{WOCS 40007}\\
\hline
2180.6428 & $-39.39\pm0.03$ & $62.76\pm0.05$ & \\
2415.8614 & $-84.48\pm0.02$ & $107.52\pm0.05$ & \\
2461.7212 & $ 80.20\pm0.02$ & $-84.10\pm0.05$ & \\
2475.6910 & $-21.15\pm0.09$ & $24.80\pm0.07$ & \\
2476.8479 & $-50.81\pm0.05$ & $61.94\pm0.04$ & \\
2856.7194 & $ 73.97\pm0.91$ & $-94.24\pm1.32$ & \\
3190.6986 & $ 15.85\pm0.03$ & $-29.40\pm0.10$ & 
\enddata
\tablenotetext{a}{mJD = BJD - 2 450 000}
\tablecomments{This table is published in its entirety in the electronic edition, but a portion is shown here for guidance regarding its form and content.}
\end{deluxetable}

Of the 41 spectra we had available, 34 covered wavelengths between
5000 and 5240 \AA, roughly centered on the Mg b triplet, 6 covered
5460 to 5730 \AA, and one covered 6233 to 6227 \AA.  The spectra in
each group were continuum normalized before being trimmed to a common
wavelength range and resampled to a common logarithmic wavelength
spacing at approximately the same pixel resolution as the original
spectra. Each of these groups was analyzed separately using a
high-resolution solar spectrum \citep{hinkle} and IDL routines
provided publicly by S. Rucinski\footnote{\tt
  http://www.astro.utoronto.ca/$\sim$rucinski/SVDcookbook.html} to
derive the broadening function via singular value decomposition. The
broadening function containing all of the singular values was then
gaussian smoothed to reduce noise --- by experimenting, we found, as
Rucinski has stated, that RVs were sometimes
systematically offset when we used a broadening function that was
composed from a limited set of the singular values.  For the reddest
spectrum, disentangling could not be used, but the broadening function
peaks for all three components were well separated.  For the bluest
group of spectra, convergence was straightforward, thanks to the large
number of spectra. For the mid-range group of 6 spectra, convergence
was more difficult because three had very similar
orbital phase ($\phi \approx 0.75$), but after restricting ourselves
initially to the three spectra most widely spaced in phase for
deriving the average spectra, convergence was achieved.

There were a handful of spectra for which two of the stars had radial
velocities that were identical to within about 2 km s$^{-1}$, and for these
disentangling failed to uniquely identify the RVs.  However, by
using the spectral disentangling technique, we were able to analyze a larger
number of spectra than we were when using broadening functions alone, and
could therefore derive a more complete phased radial-velocity curve.  We fit
gaussian profiles to the broadening function peaks that could be resolved in
each disentangled spectrum's broadening function. The central values were used
to measure the stellar RVs, after corrections for fiber velocity
offsets and heliocentric velocity.

In the case of the two eclipsing binary components (WOCS 24009 B
and C), the stars are shown to be extremely similar in their characteristics,
and so the ratio of the integrated areas of the gaussians gives us a measure
of their luminosity ratio.  We restricted ourselves to spectra in which three
component peaks could be separately measured in the broadening function of the
spectra before disentangling. This limited us to 15 of the spectra centered on
the Mg b triplet, and 5 of the spectra centered on about 5595 \AA. For these
two groups, we found $(L_C / L_B)_{\lambda} = 0.863 \pm 0.033$ and $0.880 \pm 0.014$
(after rejection of one outlier), respectively, where the error in the mean is
quoted. Because these two measurements are consistent within the
uncertainties, we combined them in a weighted mean to get a final value of
$0.877 \pm 0.013$ and we treated this as an external constraint on the
$V$-band luminosities of the two stars in the binary star models.

The temperature difference between the bright component A and the
eclipsing stars is much larger ($400-500$ K, versus less than 100 K for
the two eclipsing stars), and so the broadening functions from limited
wavelength ranges do not give a direct measure of luminosity
ratios. However, we originally thought that this might be our only
constraint on the photometric properties of component A, though this
ended up not being the case 
(see section \ref{w24009}). Using the same
spectra with resolved peaks, we estimated $l_3=L_A / (L_A + L_B +
L_C)$ using the areas of the broadening function peaks for all three
components. In the bluer region around the Mg b triplet we found
$0.593\pm0.014$, and from the region around 5595 {\AA} we found
$0.534\pm0.027$. The two values differ more than they should if they
were truly reflecting the light ratios in the two wavelength bands (as
indicated by integrations of \citealt{coelho} synthetic spectra).
We adopted a middle value for the ratio of the $V$
luminosities ($0.55\pm0.02$) as a loose constraint on the effects
that component A's light has on reducing the measured depths of the
eclipses, but the ratio was allowed to vary.

\begin{figure}[htbp]
\centerline{\includegraphics [width=0.55\textwidth]{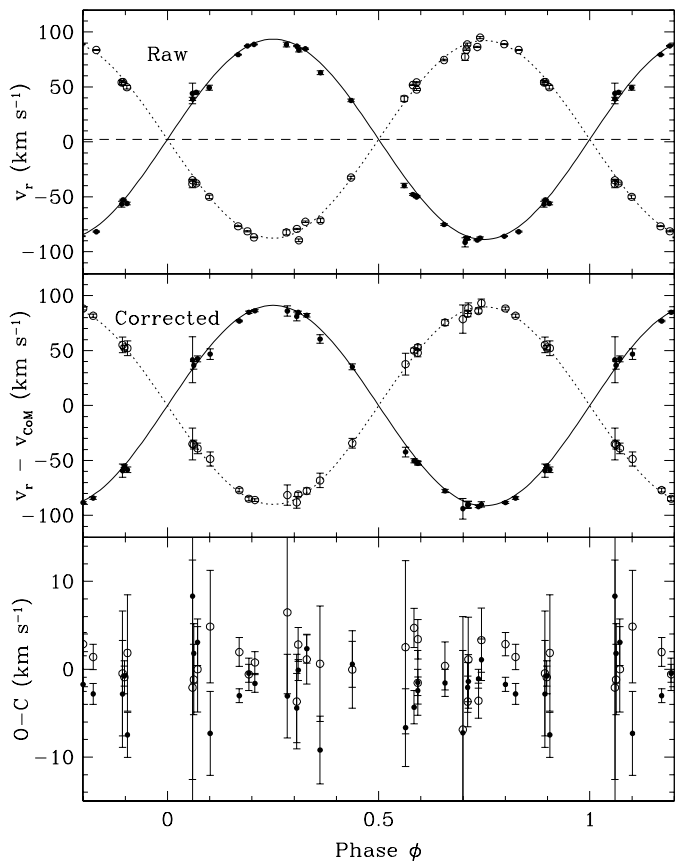}}
\caption{{\it Top panel:} Raw RVs for WOCS 24009 phased to the orbital
  period. Model curves for the two stars are shown with solid and
  dotted lines.  {\it Middle panel:} RVs with corrections for binary
  center of mass motion and for light travel time effects on the true
  orbital phase.  {\it Bottom panel:} Corrected $O-C$ diagram for the
  eclipsing binary components. For all panels, open circles are
  observations of the brighter eclipsing star and filled circles are for the fainter star.
  \label{rv}}
\end{figure}

For spectra of the WOCS 40007 system, we were able to improve the precision of
the measurements presented by \citet{jeffries13}, and also make use of some
spectra in which the lines of the eclipsing stars were fairly strongly
blended. This allowed us to measure RVs near eclipse phases, but
more importantly, it gave us new measurements of the center-of-mass velocity
for the eclipsing binary that could be used to better determine the orbit of
the faint non-eclipsing star (component C) in the system. Lines of component C could not be
identified, so we decomposed the spectra into two
components. Of the 40 available spectra, 34 were centered on the Mg b triplet,
4 covered 5460 to 5730 \AA, and two covered 6233 to 6227 \AA. There was not
enough phase coverage to apply the disentangling method to the two smaller
groups of spectra, but there was only one spectrum (in the red group) for
which the broadening functions overlapped too strongly to separate.

\noindent
\begin{figure}[!h]
\centerline{\includegraphics [width=0.55\textwidth]{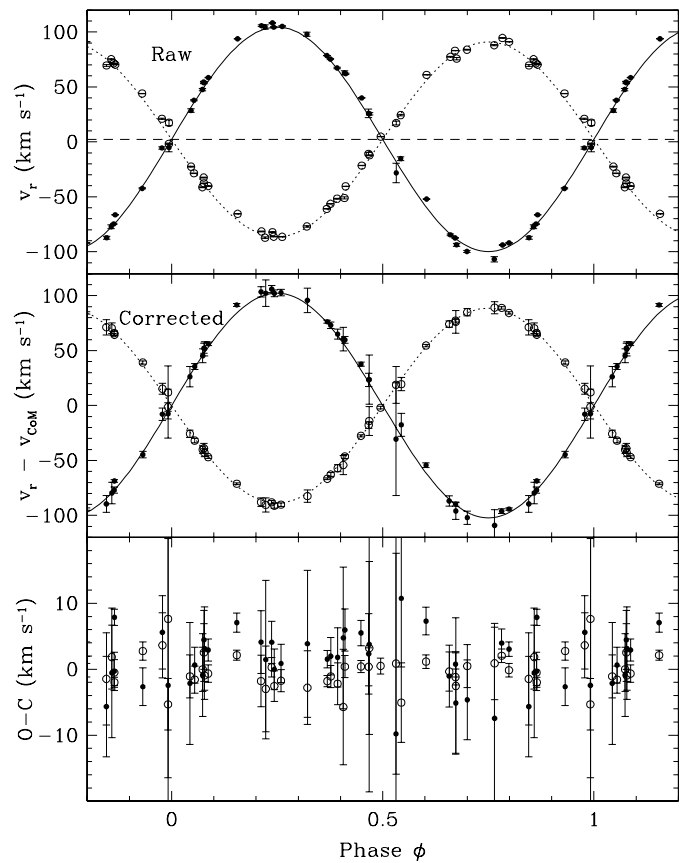}}
\caption{RVs for WOCS 40007 phased to the orbital period. In the
  middle panel, center-of-mass motion and light travel time
  corrections are applied to the RVs and phases as in Fig. \ref{rv}. For all
  panels, open circles are observations of the brighter eclipsing star and filled circles are
  for the fainter star.
  \label{mj_rv}}
\end{figure}
Table \ref{rvtab} presents the measured RVs for both systems and
Figures \ref{rv} and \ref{mj_rv} show the phased RVs for WOCS
24009 and WOCS 40007, respectively, plotted over the binary orbits.

\subsubsection{Photometric Temperatures}\label{temps}

We derived temperature estimates for the stars from the $BV$ photometry of
\citet{kalirai01} and the $VI_C$ photometry of \citet{yang}. These datasets
were selected based on their high signal-to-noise observations. Comparisons with our own
photometry indicated that the $B-V$ and $V-I_C$ colors in these datasets match
our own calibration within about 0.005 mag even though the differences in
individual wavelength bands were considerably larger. Therefore, we are
reasonably certain that the color zeropoints are reliable. After correcting
the colors for differential reddening using the reddening map of
\citet{platais13} and a mean cluster reddening [$E(B-V)=0.16\pm0.007$;
\citealt{att}]\footnote{We re-examined our earlier estimation of the cluster
  reddening from clump stars \citep{jeffries13} in light of the critique of
  recent measurements by \citet{att}. We applied differential reddening
  corrections to our clump stars, and found that this did not change the
  derived reddening. However, the corrections we made to account for
  metallicity-dependent effects on the luminosities of the clump giants in NGC 6819 relative to those in M67 are
  wavelength dependent, and modifying those result in a significant
  change. If we assume that the difference in [Fe/H] between NGC 6819 and M67 goes from $+0.09$ (with NGC
  6819 more metal rich) to 0, our derived reddening goes from $E(B-V)=0.12$ to
  0.15. While this is not proof that the larger value is correct, it does
  show that the different estimations of reddening may be consistent. If the metallicity
  difference between M67 and NGC 6819 is established to be closer to 0, the
  higher reddening value should be preferred.},
we applied the color-temperature relations of \citet{casa} and
averaged the temperatures derived from the $B-V$ and $V-I_C$
colors. The estimates from the two colors for each star generally
agreed well with each other, although the temperatures derived from $B-V$ were about 100
K hotter on average. If we assume that [Fe/H] is subsolar, the two estimates
come into better agreement.

\begin{deluxetable*}{lcccccc}[!htb]
\tablewidth{0pt}
\tablecaption{Measured Parameters for Stars in NGC 6819\label{summary}}
\tablehead{
\colhead{Star ID} & \colhead{$V$\tablenotemark{a}} & \colhead{$B-V$\tablenotemark{a}} & \colhead{$T_{\rm eff}$ (K)} & \colhead{$M / M_\sun$} & \colhead{$R / R_\sun$} & \colhead{$(m-M)_V$} }
\startdata
WOCS 23009 A & 15.130 & 0.632 & 6315 & $1.464 \pm 0.011$\tablenotemark{b} & $2.149 \pm 0.009$ & $12.39\pm0.07$\\
WOCS 40007 A & 16.111 & 0.642 & 6350 & $1.218 \pm 0.008$ & $1.367 \pm 0.003$ & $12.40\pm0.08$ \\
WOCS 40007 B & 16.949 & 0.692 & 5900 & $1.068 \pm 0.007$ & $1.090 \pm  0.002$ & $12.40\pm0.08$ \\
WOCS 24009   & 15.214 & 0.661 & & & \\
WOCS 24009 A & 15.743 & 0.589 & & $1.251\pm0.057$ & & \\
WOCS 24009 B & 16.835 & 0.721 & 5925 & $1.090\pm0.010$ & $1.095\pm0.007$ & $12.37\pm0.07$\\
WOCS 24009 C & 16.965 & 0.744 & 5855 & $1.075\pm0.013$ & $1.057\pm0.008$ & $12.37\pm0.07$
\enddata
\tablenotetext{a}{The quoted photometry is from \citet{kalirai01}, but has not been corrected for zeropoint differences in calibration or for differential reddening.}
\tablenotetext{b}{Based on a photometry-based extrapolation from other eclipsing stars in the cluster.}
\end{deluxetable*}

The temperatures are given in Table \ref{summary}. 
The stars fall into two groups with similar temperatures. The brightest star in 
WOCS 23009 \citep[hereafter WOCS 23009 A;][]{sandquist13}
and the brightest star in the eclipsing binary WOCS 40007 (hereafter WOCS 40007 A)
are on opposite sides of the turnoff with similar colors but
quite different luminosities. The stars WOCS 24009 B and C and probably WOCS 40007 B 
are also quite similar to each other due to their nearly identical masses.\\

\section{Analysis}

\subsection{Cluster Membership}\label{memb}

The information extracted from models of WOCS 24009 will only be of
use for assessing the properties of NGC 6819 if WOCS 24009 is a member
of the cluster. There are three ways in which cluster membership can
be examined.  First, the most restrictive evidence for cluster
membership comes from proper-motion measurements \citep{platais13}.
The proper motions are derived from a combination of old
photographic plates with CCD observations obtained from the 3.6 m
Canada-France-Hawaii Telescope (CFHT).
For WOCS 24009 \citep[star 592928 in][]{platais13},
the formal uncertainty of proper motion in either axis is
slightly less than 0.2 mas~yr$^{-1}$ and the star's offset proper motion from
the cluster mean motion is also $\sim $0.2 mas~yr$^{-1}$. These parameters
yield an astrometric membership probability of 99\%. 
We can also compare the systemic radial velocity to those measured for
other stars in the field to determine membership. \citet{milliman} have done an extensive radial
velocity survey of stars in the NGC 6819 field, and they use
one-dimensional Gaussian fits to the field $F_{f}(v)$ and cluster
$F_{c}(v)$ velocity distributions to compute a membership probability:
\[ p(v) = \frac{F_{c}(v)}{F_{f}(v)+F_{c}(v)} \]
Ideally, we would calculate this from the system velocity ($\gamma$)
derived from a three-body solution to the stellar orbits (see \S
\ref{tertorbit}). However, the system has not been observed over a
large enough fraction of the outer orbit period to properly determine
$\gamma$.  The velocity at which component A and the binary
center-of-mass velocities cross is approximately 2 km s$^{-1}$ though,
consistent with the cluster mean RV of 2.45 km s$^{-1}$
\citep{milliman}. This also supports membership.  The last method we
considered uses the position of the system in relation to the core
radius of the cluster, where $r_{c}$ is a King model value \citep[1.8
  $r_{c}$;][]{kalirai01}. A membership probability is estimated as
45\% using a ratio of the spatial density of probable cluster members
to the total spatial density, but this method only weakly constrains
the membership because it does not take into account differences in
the cluster and field distributions in the CMD.  Other evidence that
supports cluster membership are the physical similarities between WOCS
24009 eclipsing binary stars and the well-measured stars in WOCS 40007
\citep{jeffries13}, for which the membership information is more
complete.  Based on the evidence available to date, we find no reason
to doubt that WOCS 24009 is a cluster member.

\subsection{Radial-Velocity and Light-Curve Modeling}

Three stellar components were identifiable in spectra taken of WOCS 24009 even
before the {\it Kepler} mission began. However, due to the similar
characteristics of the fainter two stars and significant blending with the
brightest component in many spectra, an orbital solution had not been
determined. The {\it Kepler} photometry not only made it possible to
differentiate between the two eclipses of very similar depth, but the eclipse
timings also showed clear evidence of a light travel time (LTT) effect resulting
from the gravitational influence of a third star on a wide orbit. It also
became clear that scatter in the RVs of the eclipsing stars was
being introduced by reflex motion of the binary in response to the motion of
component A. The eclipse timings are given in Table \ref{etimes} with
the earliest eclipse observed from the ground in July 2001 \citep{talam10}.

We therefore sought a consistent solution assuming that the system is
hierarchical: the eclipsing stars in the binary follow Keplerian
orbits, and component A and the barycenter of the binary also
follow much larger Keplerian orbits. The modeling of the stars was
therefore separated into two parts: characterization of the radial
velocity curve of the eclipsing binary and the wide orbit of the
component A using the RVs and eclipse timings, and
characterization of the light curves to derive final masses and radii
for the eclipsing stars.

\begin{figure*}[!ht]
\centering
\vspace{-40pt}
\includegraphics [width=0.8\textwidth]{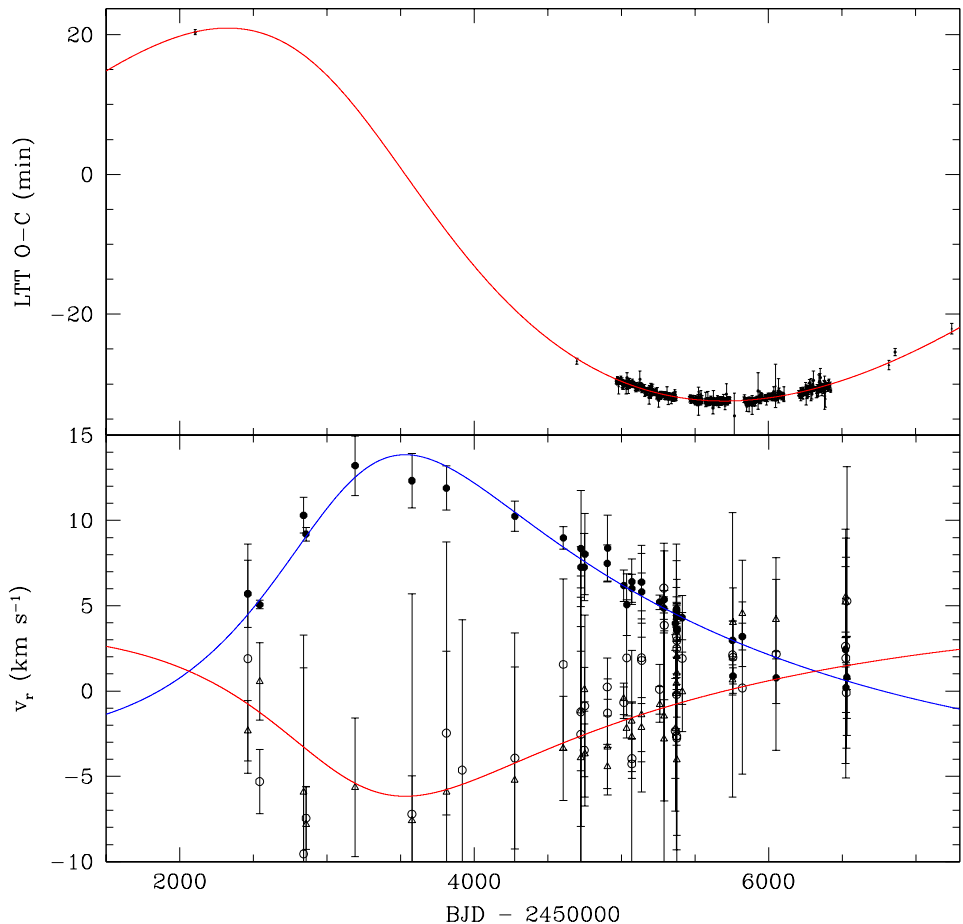}
\vspace{-100pt}
\caption{Orbit modeling solutions for the third star in WOCS 24009, plotted to fit the RVs and eclipse timings
as described in \S \ref{tertorbit}.
{\it Top panel:} $O-C$ for variations in binary eclipse timings. {\it
    Bottom panel:} Observed RVs of the bright third star WOCS 24009 A
  ({\it solid}) and binary star components ({\it open}) after subtracting
  binary star model predictions. Binary star model curves are shown for the third star ({\it blue}) and binary
  barycenter ({\it red}).
  \label{tertfig}}
\end{figure*}

Models for the photometric and radial-velocity measurements were
computed using the ELC code \citep{orosz00}. ELC uses a $\chi^{2}$
statistic to determine the goodness-of-fit where the overall
$\chi^{2}$ is the sum of the $\chi^{2}$ for each photometry filter,
RVs for each binary component, and any observational
constraints. Genetic and Markov chain optimizers were used to
determine best-fit models in the multi-dimensional parameter space. We
generally began with the genetic algorithm in order to search large
volumes of parameter space to arrive at the best fit minimum for each
binary parameter. After the initial identification of a best fit, the
uncertainties on individual photometric and radial-velocity
observations were scaled in order to produce a reduced $\chi^{2}$ of
1. This is essentially equivalent to replacing the a priori
uncertainty estimates with a posteriori estimates that can account for
random systematic effects. Subsequently, we used Markov chain Monte
Carlo modeling \citep{tegmark04} to determine the solution and search
alternative models near the parameter minima to estimate parameter
uncertainties. $1\sigma$ uncertainties were found by identifying the
parameter ranges covered by solutions that had a $\chi^{2}$ (not
reduced $\chi^2$) value within 1 of the minimum \citep{avni76}.

\begin{deluxetable*}{lcccc}[!ht]
\tablewidth{0pt}
\tablecaption{Orbit Solution for the Binary Stars in WOCS 24009 and WOCS 40007\label{tert}}
\tablehead{\colhead{Parameter} & \multicolumn{2}{c}{WOCS 24009} & \multicolumn{2}{c}{WOCS 40007}\\
& This Study & Milliman et al. & This Study & Milliman et al.}
\startdata
$P_{bin}$ (d) & $3.6493010\pm0.00000018$ & 3.64921$\pm$0.00004 & 3.18509586$\pm$0.00000008 & $3.185087\pm0.000012$\\
$P_3$ (d) & $8333\pm30$ & & $3248\pm3$ & \\
$t_{bin}$ (HJD - 2450000) & $2104.85489\pm0.00017$ & 4836.1$\pm$1.4 & $2102.88690\pm0.00004$ & $4329.3\pm2.4$ \\
$t_3$ (HJD - 2450000) & $3182\pm13$ & & $5601.3\pm2.5$ & \\
$e_3$ & $0.368\pm0.003$ & & $0.307\pm0.002$ & \\
$i_3$ ($^{\circ}$) & 68.7 & & & \\
$\omega_3$ ($^{\circ}$) & $145.0\pm0.6$ & & $72.4\pm0.3$ & \\
$K_{bin}$ (km s$^{-1}$) & $4.74\pm0.02$ & & $4.764\pm0.009$ & \\
$K_A$ (km s$^{-1}$) & $8.21\pm0.35$ & & $88.9\pm0.4$ & $87.9\pm1.0$ \\
$K_B$ (km s$^{-1}$) & $88.86\pm0.51$ & $88.0\pm1.3$ & $101.6\pm0.3$ & $103.3\pm1.1$ \\
$K_C$ (km s$^{-1}$) & $90.05\pm0.35$ & $90.0\pm1.3$ & & \\
$q_{bin}=M_2/M_1$ & $0.987\pm0.007$ & $1.023\pm0.023$ & $0.875\pm0.005$ & $0.851\pm0.014$\\
$q_A=M_A/(M_B+M_C)$ & $0.578\pm0.026$ & & & \\
$\gamma$ (km s$^{-1}$) & $3.17\pm0.12$ & $0.4\pm0.7$ & $2.83\pm0.18$ & $3.7\pm0.5$
\enddata
\end{deluxetable*}

\subsubsection{Orbit Modeling of Third Star}\label{tertorbit}

When modeling the RVs and eclipse timings for WOCS 24009, we
conducted simultaneous fits for the following parameters: period $P_{bin}$ 
and time of primary  eclipse $t_{bin}$ for the eclipsing binary 
(in its comoving reference frame);  period $P_A$, time of periastron 
$t_A$, eccentricity $e_A$, and argument of periastron $\omega_A$ for 
the third star's orbit; the radial-velocity amplitudes for the third star $K_A$ 
and the stars in the eclipsing binary ($K_B$ and $K_C$), and the radial 
velocity of the barycenter of the system $\gamma$. However, it became 
clear that some aspects of the third star's orbit are not well constrained with 
the current data. The basic problem is that, even though we have data 
covering approximately 4000 d, by chance we have observed component A
orbit essentially between times of zero radial velocity with one velocity 
maximum (closest to periastron) in between. Because of the distribution of the
observations, there are correlations between the system velocity, the
velocity semi-amplitudes of the third star $K_A$ and binary $K_{bin}$, and
the eccentricity.

Because the indications are that the system is a
member of NGC 6819, we fixed the system velocity to the cluster
average \citep{milliman}. This choice may introduce systematic errors
into the values of $K_A$ and $K_B$, but there are good reasons for trying
to model the third star's orbit as best we can right now.
First, we can correct the binary star velocities to first order for the motion 
induced by component A, producing more precise measurements of the velocity
amplitudes $K_B$ and $K_C$, and thus the masses of the stars. The radial
velocities of the eclipsing stars are directly affected by the
center-of-mass motion, but light travel time also affects the inferred phase
of each observation. This effect is relatively important in this system (at
the 0.01 level in phase) because of the large size of the third star's orbit and
the relatively short period of the eclipsing binary. The top two panels
of Figure \ref{rv} illustrate the radial-velocity variations with and without
the corrections to the RVs and phases of each observation.  Second, the ratio 
of the velocity amplitudes allow us to
determine the mass ratio $M_A / (M_B + M_C)$, and in principle, the mass of
the third star. The size of the third star cannot be determined because
there is no evidence that component A eclipses either binary component in {\it
  Kepler} photometry. However, our photometric decomposition indicates
that the third star is the one closest to the cluster turnoff with a 
{\it measurable} mass, in spite of the lack of eclipses.

Figure \ref{tertfig} shows the results of our best preliminary fit to the eclipse
timings and RVs. The top panel shows the variations in the
binary star eclipse timings compared to a linear ephemeris (observed minus
computed: $O-C$) over time. The error bars are scaled to give a
reduced $\chi^{2}$ of about 1 for the RVs and LTT separately. This was
needed to ensure that neither dataset was weighted too heavily in finding a
joint solution. The bottom panel shows observed radial
velocities of the binary center of mass and the third star. 

To obtain this fit, we had to apply some constraints based on an analysis of
the observations. Extrema in the top plot of the eclipse timings correspond to
times when the binary center-of-mass velocity crosses the barycenter velocity
$\gamma$, and these are well constrained by the radial-velocity
observations. The first several radial-velocity observations (BJD before
2453000) indicate that the crossing occurred around BJD 2452100-2452300, very
close to the first ground-based eclipse observation (BJD 2452105). Similarly,
the observations by {\it Kepler} have covered the LTT minimum (near BJD
2455900), as corroborated by the RVs. As a result, the full LTT
amplitude is constrained to be close to 56 minutes.  The eccentricity of the
outer orbit is clearly visible in the asymmetry of the RVs
between the zero crossings. Note that the radial-velocity curve for the
eclipsing binary is related to the derivative of the eclipse timing curve,
with the radial-velocity extrema occurring at inflection points in the LTT
curve.  Unfortunately, the period of the third star's orbit will not be
constrained more tightly until more eclipse times are measured as the
eclipsing binary moves more rapidly away from us. We are also unable to
directly measure the radial-velocity amplitudes because both velocity extrema
have not been observed and because the precise system velocity for the triple
is not well constrained yet.

The best current fit for the eclipsing binary orbit is given in Table \ref{tert}.  For
comparison, we include the radial-velocity fits from \citet{milliman}.  We
expect our results to be more precise and accurate on account of the improvements in
technique for deriving the RVs from the complex spectra and,
more importantly, due to our control for the effects of the third star. In the
important case of the velocity semi-amplitudes for the eclipsing stars $K_{B,C}$, a
comparison shows that there are systematic differences in values by
approximately $1\sigma$ compared to the Milliman et al. fits, and a significant
decrease in uncertainty by a factor of $2-3$. The lower $K$ uncertainties
translate to a corresponding reduction in the uncertainties in the derived masses.
With the fit for the third star's orbit in hand, we can compensate for its effects on the
binary radial-velocity measurements. These values are used as constraints in
our models of the eclipsing binary light curve (see \S \ref{ELC}). Using the
mass ratio from this analysis ($q_{bin}=0.987\pm0.007$), we can determine
center-of-mass velocities for the eclipsing binary at each epoch, and use them
to measure the mass ratio $q_A = M_A / (M_B+M_C)$. Using the
weighted average of the eight best measurements near the velocity extremum
(BJDs 2452461 to 2454747), we derive a value $q_A = 0.578 \pm 0.026$.
 
 \begin{figure*}[!ht]
\centering
\includegraphics [width=.9\textwidth]{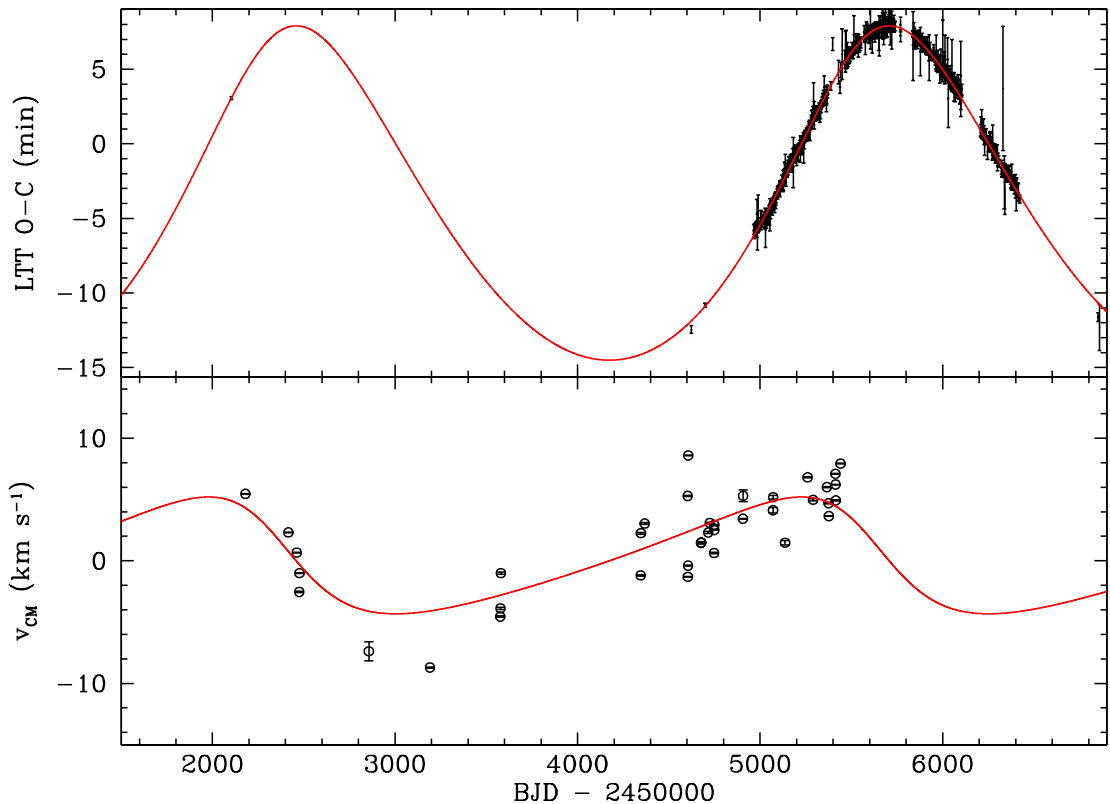}
\vspace{-20pt}
\caption{Orbit solution for the third star of WOCS 40007, similar to Fig. \ref{tertfig}.
{\it Top panel:} $O-C$ for variations in binary eclipse timings. {\it
    Bottom panel:} Observed RVs of the eclipsing binary 
    center of mass compared with model predictions.
  \label{mjtertfig}}
\end{figure*}
 
We calculate the mass of the third star (WOCS 24009 A) to be $M_{A} =1.251 \pm
0.057 M_{\sun}$.  The inclination of its orbit can be estimated using
Kepler's third law and the masses and period, either by comparing the expected
radial-velocity amplitudes to the measured ones or by comparing the expected
size of the orbit to the amplitude of the light travel time effect on the
eclipses (which is related to the projection of the orbit along the line of
sight). At present, the light travel time amplitude is much better constrained
than the radial velocity semi-amplitudes for the binary $K_{bin}$ and third star
$K_{A}$. So using the LTT results, we estimate the inclination to be
$i_A=68$\degr. Because the inclination is quite far from edge-on, the third star
will not eclipse and its radius cannot be measured directly. The most
important information it will provide will come from its mass and good
photometry. Three-body (non-Keplerian) dynamics could affect how precisely we
are able to determine the characteristics of the third star's orbit,
especially when all of our observations span approximately one orbital
period. However, some exploratory calculations (D. Short, private
communication) indicate that the effects are likely to be quite small as a
result of the large difference in orbital periods.  The difficulties involved
in measuring the motions of the binary center of mass and the third star means that
the third star's mass will not be a strong constraint on cluster properties at
present. Future monitoring should bring down the mass uncertainty
significantly.

Finally, because additional {\it Kepler} observations were available and
because we improved our analysis of the radial-velocity measurements for the
WOCS 40007 binary \citep{jeffries13}, we have updated our models of that
triple system. Its third star is considerably fainter than the eclipsing
binary, so that we are unable to measure its RVs from spectra and
will not be able to derive its mass. However, improved models of the third star's
orbit will allow us to reduce the mass uncertainties for the eclipsing
stars. The combination of RVs and eclipse timings has given us
observations covering more than one complete orbit of the third star, although
with differing degrees of precision. The improved RVs more
clearly draw out the shape of the velocity curve, and have significantly
increased the measured eccentricity. The additional {\it Kepler} eclipse
timings have allowed us to precisely identify both inflection points in the
LTT O-C diagram, which in turn determines the duration of the radial-velocity
swing from maximum to minimum. The velocity semi-amplitudes for the eclipsing
binary have not changed greatly since our original determination, but the
uncertainties have improved by about a factor of two. An updated plot of
the third star's orbital effects is shown in Figure \ref{mjtertfig}.

\subsubsection{Binary Star Modeling}\label{ELC}

\begin{deluxetable*}{lccc}[!h]
\tabletypesize{\footnotesize}
\tablewidth{0pt}
\tablecaption{Orbital Solutions: Limb-Darkening Law\label{orb}}
\tablehead{\colhead{Parameter} & \colhead{Kepler + Ground} & 
\colhead{Kepler Only} & \colhead{Model Atmospheres}}
\startdata
\vspace{0.05in}
Constrained: & & &\\
$t_o$ (HJD - 2450000)&\multicolumn{3}{c}{2104.8424 $\pm$ 0.0002}\\
$P$ (d)&\multicolumn{3}{c}{3.64930279 $\pm$ 0.00000016}\\
$q = M_C/M_B$&\multicolumn{3}{c}{0.987 $\pm$ 0.007}\\
$K_B$ (km s$^{-1}$) & \multicolumn{3}{c}{$88.86 \pm 0.54$}\\
$L_{C}/L_{B}$ & \multicolumn{3}{c}{$0.877 \pm 0.013$}\\
$T_A$ (K) & \multicolumn{3}{c}{$6330 \pm 53$} \\
$T_B$ (K) & \multicolumn{3}{c}{} \\ 
\hline
Fitted: & & & \\
$i$ ($^{\circ}$) & 89.27 $\pm$ 0.02 & 89.28 $\pm$ 0.02 & 89.42 $\pm$ 0.02\\
$R_B/a$ & 0.0853 $\pm$ 0.0002 & 0.0853 $\pm$ 0.0001 & 0.0856 $\pm$ 0.0002\\
$R_B/R_C$ & 1.029 $\pm$ 0.003 & 1.025 $\pm$ 0.003 & 1.020 $\pm$ 0.003\\
$T_C/T_B$ & 0.988 $\pm$ 0.0002 & 0.988 $\pm$ 0.0002 & 0.989 $\pm$ 0.0002\\
$R_A^{2}/R_B^{B}$ & 2.126 $\pm$ 0.024 & & 2.071 $\pm$ 0.023\\
Contam & 0.047 $\pm$ 0.003 & 0.070 $\pm$ 0.003 & 0.053 $\pm$ 0.003\\
\hline
Derived: & & & \\
$M_B$ ($M_{\sun}$) & 1.089 $\pm$ 0.018 & 1.090 $\pm$ 0.023 & 1.090 $\pm$ 0.019\\
$M_C$ ($M_{\sun}$) & 1.076 $\pm$ 0.016 & 1.076 $\pm$ 0.021 & 1.076 $\pm$ 0.018\\
$R_B$ ($R_{\sun}$) & 1.099 $\pm$ 0.006 & 1.100 $\pm$ 0.007 & 1.104 $\pm$ 0.007\\
$R_C$ ($R_{\sun}$) & 1.069 $\pm$ 0.006 $\pm$ 0.013\tablenotemark{a} & 1.073 $\pm$ 0.007 & 1.083 $\pm$ 0.007\\
$R_C/a$ & 0.0829 $\pm$ 0.0002 & 0.0832 $\pm$ 0.0002 & 0.0839 $\pm$ 0.0001\\
$(R_B+R_C)/a$ & 0.1682 $\pm$ 0.0002 & 0.1684 $\pm$ 0.0002 & 0.1695 $\pm$ 0.0001\\
$a$ ($R_{\sun}$) & 12.90 $\pm$ 0.07 & 12.90 $\pm$ 0.08 & 12.90 $\pm$ 0.08\\
log $g_B$ (cgs) & 4.392 $\pm$ 0.003 & 4.392 $\pm$ 0.003 & 4.389 $\pm$ 0.003\\
log $g_C$ (cgs) & 4.411 $\pm$ 0.003 & 4.408 $\pm$ 0.003 & 4.400 $\pm$ 0.003
\enddata
\tablenotetext{a}{The two uncertainties are estimates of the random and systematic effects.}
\end{deluxetable*}

Before we started to fit the light curves with binary star models, we
trimmed the photometric data to observations taken in and near
eclipses. {\it Kepler} photometry shows that there is negligible
variation outside of eclipse, indicating that ellipsoidal variations
(due to non-spherical stars) and reflection effects can be
ignored. Variations in the out-of-eclipse photometry (due to
systematics in the {\it Kepler} photometry or atmospheric effects in
the ground-based photometry) added unconstructively to the $\chi^2$ of
our fits, and substantially lengthened the time needed to compute model fits, so we
expunged these points. We also applied zeropoint offset corrections to nights of
ground-based data, where the offset was determined from the median
magnitude of any out-of-eclipse data from that night.  These
corrections were made to bring the ground-based light curves into
greater consistency, and are likely to be the result of stellar
activity. Most of the nights only had out-of-eclipse photometry
available either from the evening or morning, so the median magnitude
was taken for that and applied to the whole night. For nights with
out-of-eclipse data on both sides of the eclipse, the median magnitude
was determined using the side with the largest number of datapoints
and least uncertainty in the magnitudes.

The effects of limb-darkening were examined by fitting the data with
two different types of models: one using a quadratic limb-darkening
law, and the other using PHOENIX model atmospheres
\citep{hauschildt97}. When using a quadratic limb-darkening law, we
computed models using {\it Kepler} data both with and without our
ground-based observations. In both types of models we fixed one
coefficient in the {\it Kepler} filter band for each star using ATLAS
atmosphere values from \cite{claret00} and fitted for the other, which
allowed ELC greater freedom to search for the coefficient values for
each filter that produce the best fits to the light curves.  By fixing
one coefficient and fitting for the other, systematic error in the
assumed coefficient (due to incorrect $T_{\rm eff}$, log $g$, or
metallicity) can be partly compensated for because coefficients tend
to be correlated.  With the more precise {\it Kepler} data, the
coefficient is better constrained than in the ground-based
data. Finally, because PHOENIX model atmospheres describe the limb
darkening of the stars and variation of emitted intensity with
emergent angle, there is no need to assume limb-darkening
coefficients. On the flip side, systematic errors in the atmosphere
models can introduce systematic errors in best-fit parameters. We have
computed models using both limb darkening descriptions as a way of
assessing the systematic uncertainties and their effects on the
measured stellar characteristics.

Results for the two types of models can be found in Table
\ref{orb}. We fit for 10 free parameters in the light curve models:
inclination $i$, ratio of the brightest eclipsing star radius to the semi-major
axis $R_B/a$, ratio of radii of the eclipsing stars $R_B/R_C$,
temperature of the brighter eclipsing star $T_B$, ratio of the
temperatures of the eclipsing stars $T_C/T_B$, temperature of the
third star $T_A$, ratio of the surface area of the bright component A to
the surface area of the brightest eclipsing star $R_A^{2}/R_B^{2}$, binary mass
ratio $q = M_C/M_B = K_B/K_C$, velocity semi-amplitude of the brightest eclipsing star $K_B$, and external
contamination of the {\it Kepler} light curve. {\it Kepler} photometry
provides a very tight constraint on the shape of the light curves, but
not their absolute depths due to the contamination by unassociated
stars projected near WOCS 24009 on the sky. Our ground-based data,
which is much less affected by contamination from unassociated stars,
provides better constraints on the absolute depths of the eclipses.

We have applied the following spectroscopic constraints to the light
curve models during the fits: temperature of the third star $T_A$, the
luminosity ratio for the eclipsing stars $L_C/L_B$, velocity
semi-amplitude $K_B$, and binary mass ratio $q = K_B/K_C$. $q$, $K_B$,
and $T_A$ are included as free parameters to allow their values to
vary within the uncertainties of the triple star fit and spectroscopic
temperature measurement in order to incorporate this uncertainty in
the overall uncertainty for the computed binary parameters.

\section{Results and Discussion}\label{results}

\begin{figure*}[!ht]
\centering
\includegraphics [angle=270,width=0.9\textwidth]{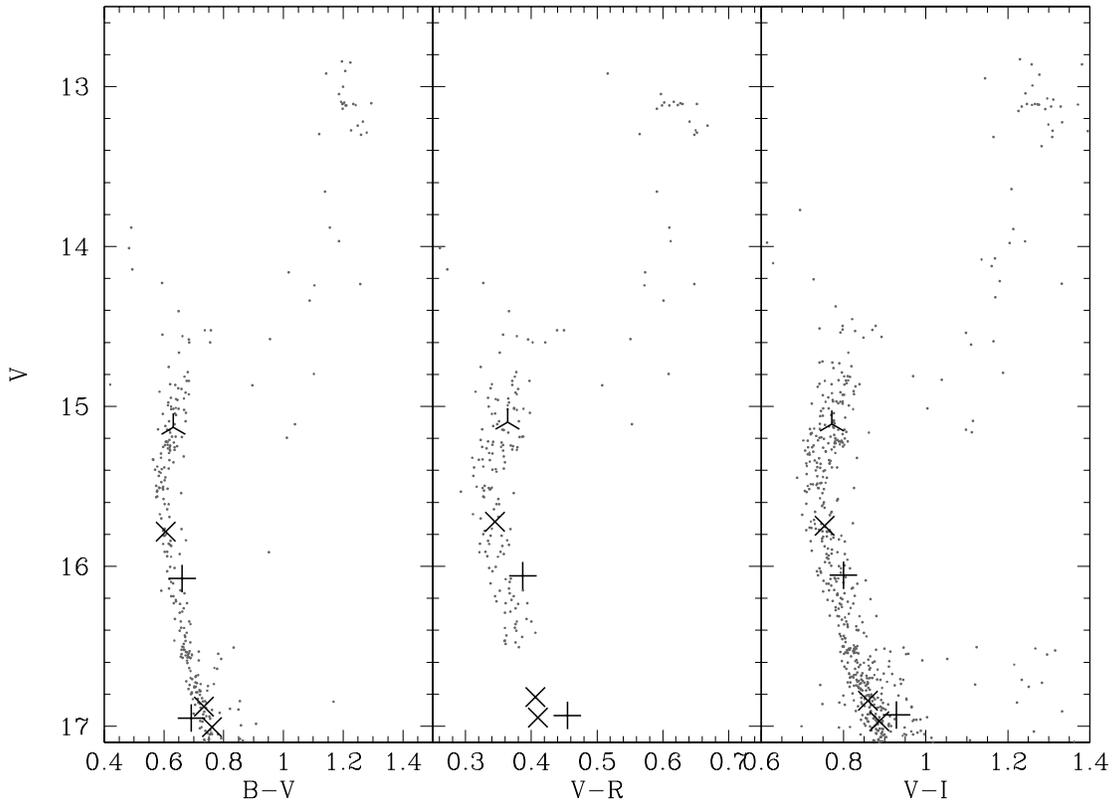}
\vspace{-20pt}
\caption{CMDs for NGC 6819 in $B-V$ \citep{kalirai01}, $V-R_c$ \citep{hole09},
  and $V-I_c$ \citep{yang}. The sample has been restricted to likely members,
  but has not been adjusted otherwise. The components of the eclipsing
  binaries WOCS 23009 (three-point star), WOCS 24009 ($\times$), and WOCS
  40007 ($+$) are shown.
  \label{cmda259}}
\end{figure*}

Before we begin a discussion of the cluster age, it is necessary to
discuss recent constraints on the cluster's metallicity. Until
recently, the only direct measurement of the composition was
\citet{bragaglia01}: [Fe/H]$=+0.09$ from three red clump
stars. However, additional measurements have become
available. \citet{att} used Str\"{o}mgren photometry to derive
[Fe/H]$=-0.06\pm0.04$. While this result does depend somewhat on the
reddening, this can be derived from the photometry for individual
stars and combined to get a robust cluster average, and the authors
did a thorough job of examining quality issues that could have
affected their zeropoints. The APOGEE experiment has also been taking
infrared spectra of a large sample of mainly red giants and red clump
stars, including observations in the {\it Kepler} field. NGC 6819 is one of
the clusters being used to calibrate stars of near solar metallicity,
and the metallicities derived from comparisons with synthetic spectra
are being corrected to match literature values. That said, APOGEE has
provided their uncalibrated abundance values, which have been lower
than their calibrated values by 0.03-0.05 dex. For example,
\citet{aspcap} derived an uncalibrated [M/H]$=+0.02$ for NGC 6819
using data from Data Release 10. We extracted measurements for 34
stars identified as asteroseismic members in \citet{stello11} from
Data Release 12, and find $+0.09$ for both the uncalibrated [M/H] and
for [Fe/H] specifically. (The calibrated value of [M/H] is 0.05 dex
higher.) APOGEE abundances are not determined truly differentially
with respect to the Sun: they use \citet{asp05} solar abundances (with
$Z_\odot=0.0122$) as a reference for producing synthetic stellar
models. More recent compilations have a higher solar abundance. For
example, if $Z_\odot = 0.0134$ \citep{asp09} was used instead, it
would result in a lower zeropoint for the APOGEE metallicity scale by
about 0.04 dex.  Finally, \citet{leebrown} presented high-dispersion
abundance analysis of a large sample (more than 240) of likely single
cluster members determined differentially with respect to solar
spectra. Their quoted values are [Fe/H]$=-0.03\pm0.06$ for the whole
sample, and their preferred value $-0.02\pm0.02$ for a subsample of
main sequence stars.

To determine a most likely range of metal contents to use in
comparisons with models, we must consider both the differences in
measured [Fe/H] and the lingering uncertainty in the solar metal
abundance $Z_\odot$. The value of $Z_\odot$ affects the APOGEE
abundance for NGC 6819 as described above, but different sets of
isochrones also assume different values. It is safest to compare
isochrones with the same $Z$, rather than [Fe/H] if we are to avoid
unnecessarily obscuring real physics effects. The values of $Z_\odot$
in recent analyses go from 0.0134 \citep{asp09} to 0.0153
\citep{caffau} --- still a significant source of uncertainty. We
believe that the best range for NGC 6819 comparisons goes from $Z =
0.012$ (using the \citealt{leebrown} measurement of [Fe/H]$\approx
-0.03$ and $Z_\odot = 0.0134$) up to $Z=0.015$ (using APOGEE
abundances for their assumed $Z_\odot$). The ratio of these
two values corresponds to approximately a 0.1 dex difference in [Fe/H].

\subsection{Interactions within the WOCS 24009 and 40007 Eclipsing Binaries}\label{inflate}

If we are going to use the binaries of NGC 6819 to constrain the
cluster age, we need to be certain that interactions have not modified
the stellar characteristics significantly. While the eclipsing binaries WOCS
24009 and WOCS 40007 are detached and have not exchanged mass, they 
are close enough that there are legitimate
concerns about the effects of tidal interactions. For well-measured
short period ($0.6 - 2.8$ d) binaries with masses in the range
$0.8-1.1M_\sun$, it has been seen that stars sometimes appear larger
by 10\%-20\% than predicted by single-star evolutionary models \citep{clausen09}.
To explain what is physically occurring inside stars with radii that
are larger than expected from theory, \citet{torres06} suggested that
synchronized rotation in short-period binaries leads to magnetic field
production and increased chromospheric activity \citep{chabrier07}.
For stars with masses of about a solar mass or less, the magnetic
activity can affect the flow of energy through the convective envelope
by reducing the efficiency of convective heat transport forcing the
star to grow larger to compensate. \citeauthor{torres06} also showed
that stars with anomalously large radii can be modeled with the same
age as their companion if a lower mixing length parameter is
used. Most longer period binaries show little or no evidence of
anomalies, but the transition range of periods is ill-defined.

The two close detached binaries we discuss in this paper have periods in the
sparse $3-4$ day range: WOCS 40007 \citep[$P$ = 3.185 d;][]{jeffries13}
and the current subject WOCS 24009 ($P$ = 3.65 d).  These binaries carry the
added benefit that they are members of a cluster with an externally
constrained age and a fairly well determined chemical composition, giving
us a better theoretical idea of what the stellar characteristics should
be. Therefore these systems can extend our understanding of the causes of
stellar inflation.

\citeauthor{jeffries13} found some indication that the radius of WOCS 40007 B
is larger than expected from models with the spectroscopic composition
\citep{bragaglia01} and preferred age of 2.5 Gyr, but at the time the
precision of the mass measurement prevented this from being a strong
statement. They considered the possibility that perhaps the adopted
metallicity or some other composition variable (most importantly, helium) was
offset from the cluster value. They examined models with decreased metallicity
that produced a slightly better consistency between the ages implied by both
eclipsing stars in WOCS 40007. Ages implied by the lower metallicity were lower by about
$\sim$ 0.2 Gyr and were more consistent with ages derived from the CMD fits.
However, with the improvement of the mass measurements for WOCS 40007, the
differences between observations and models are considerably more significant,
and radius inflation may be necessary to explain the characteristics of component B, 
and possibly both stars.

The eclipsing stars of WOCS 24009 provide an empirical basis for judging the
radius of WOCS 40007 B --- the masses of the stars are similar (so
they should have comparable internal stellar structure), and they were
formed in the same cluster (so that their initial chemical
compositions should be identical). However, WOCS 24009 has a longer
period, and should show somewhat weaker evidence of
rotationally-induced activity. Indeed, we find that the radii of WOCS
40007 B and WOCS 24009 B and C are very similar even though WOCS 40007
B is $0.02-0.03 M_\sun$ smaller in mass. So there is some indication
that the radius of WOCS 40007 B is inflated by a few
percent.

Results for other close binaries provide 
additional perspective. Detached systems with solar-type stars and periods
shorter than 3 d (e.g. EF Aqr; \citealt{vos12}) are generally circularized,
leading to rapid rotation ($v \sin i =20-73$ km s$^{-1}$) synchronized with the
orbit. These binaries generally have at least one component with a radius
measurement about 10\% larger than expected from models, show signs of
induced magnetic activity in the form of X-rays, have Ca II H and K lines in
emission, and light variations attributed to starspots. On the other hand,
binaries with periods as short as 6.94 d (EW Ori, \citealt{clausen10}) show
good agreement with standard models of solar-type stars. One other system with
a period between 3 and 7 d contains a lower-mass star with
significant radius inflation (V636 Cen: $P = 4.3$ d, $M_2 = 0.854\pm0.003
M_\sun$; \citealt{clausen09}), but the orbits in that system are significantly
eccentric ($e=0.135$) and the stars rotate pseudosynchronously with the orbit.

The larger period of WOCS 24009 compared to WOCS 40007 probably has a
relatively small effect itself --- it probably means that the rotational
velocities of the stars in WOCS 24009 are lower by about 15\%. If the WOCS
24009 stars are rotating synchronously with their orbit, the equatorial rotation
speed should be about 15 km s$^{-1}$. Probably more important is the larger
mass of the WOCS 24009 stars. Even though the mass difference is small, this
means that the convective envelopes are less massive by as much as a factor of
2 \citep[for example, see Fig. 16 of][]{torres06} between WOCS 40007 B and
WOCS 23009 A, and the radial extent of the convective zones are reduced by an
even greater factor.  These differences might be expected to reduce the
effects of magnetic activity more drastically.

The photometry of WOCS 40007 B can be decomposed
accurately from the other two stars thanks to the total eclipses. Using the
eclipse depths we derived in $BVR_cI_c$, we can put component B in
different CMDs, as seen in Figure \ref{cmda259}. When $B-V$ color is used, 
component B falls slightly to the blue of the main sequence, but in $V-R_c$
and $V-I_c$ colors, component B falls significantly to the red of the main
sequence. This also holds true for a comparison with the WOCS 24009 eclipsing
stars as well. Checking the dates of the calibrating observations for the
sources (\citealt{kalirai01} for $B-V$, \citealt{hole09} for $V-R_c$, and
\citealt{yang} for $V-I_c$), we see no reason to believe that the CMD
photometry for WOCS 40007 was affected by eclipses. According to the reddening
map from \citet{platais13}, WOCS 40007 is in a slightly more heavily reddened
portion of the cluster field. This can partly explain the red positions of
WOCS 40007 B in $V-R_c$ and $V-I_c$, but we do not have an explanation for the
bluer position in $B-V$.

A remaining question is whether the even more massive star WOCS 40007 A has
had its radius affected. When the more massive star in a short period detached
binary is expected to have a very low mass convective envelope, the radius
generally agrees better with standard stellar models (e.g.  FL Lyr,
\citealt{popper86}; V1061 Cyg, \citealt{torres06}).  However, in field
binaries, modest increases in radius can be disguised as larger ages if the
age is not otherwise constrained, as it is in clusters (see V375 Cep in the
cluster NGC 7142; \citealt{sandquist13b}). The convective envelope of WOCS
40007 A is theoretically expected to be less massive than those in the WOCS
24009 stars by another factor of more than 2. A combination of somewhat larger
than average reddening for WOCS 40007 and a correction for the contribution of the
faint third star is able to shift the brighter eclipsing star into the middle of the main
sequence band. Because the three stars appear to share the same mass-radius
isochrone, we will treat WOCS 40007 A as probably unaffected by the
interaction with its companion.

To further study the connection between chromospheric activity and
stellar size, more examples of well-studied binaries in the suggested
mass range with accurately measured mass, radius, and temperature will
be needed to map out the dependence of radius inflation on orbital
period/rotational velocity and stellar mass/convection zone depth. In
the sample of eclipsing stars in NGC 6819, component B in WOCS
40007 shows the strongest evidence of tidal inflation as a result of
the interaction with its companion component A, but additional work on the
eclipsing binaries will be needed to improve the significance of the
comparison.

\subsection{Photometry of the Component Stars}\label{debphot}

Part of the effort in this paper has been to incorporate all of our
results for the eclipsing binaries of NGC 6819 into a better overall
constraint on the characteristics of the star cluster.  Because the
photometry of individual stars (in conjunction with their masses)
contains precision information that can contribute to age
determination, we discuss our procedure for decomposing the light of
the different multiple star systems into each component.

\subsubsection{WOCS 40007}

The total eclipse of WOCS 40007 B gives us a
precise means of separating the light of component B. We remeasured the
eclipse depths after the re-reduction of the ground-based photometry described
in \S \ref{phot}. We only included nights for which we could measure both the
eclipse minimum and out-of-eclipse levels. The results are shown in Table
\ref{a259depths}, and the resulting magnitudes for component B are shown in Table
\ref{summary}. The light of component A and the faint third star cannot be as precisely
disentangled. We drew magnitudes for the third star from an isochrone
fitting the cluster's CMD and found a value that put the brightest eclipsing star
near the isochrone. This assumption might not be correct if tidal
interactions with component B result in significant changes to the
star's temperature. However, based on the results from other close binaries,
we expect stars with less massive surface convection zones (like WOCS 40007 A)
to be less affected, as discussed in \S \ref{inflate}.

\subsubsection{WOCS 24009}\label{w24009}

On the face of it, we might seem to have limited constraints on the
distribution of light among the three stars in WOCS 24009. The eclipses
of almost equal depth corroborate the idea that the eclipsing stars are nearly
identical. However, the precision of the
photometry of component A is limited by our ability to disentangle its light
from that of the eclipsing binary. Component A's photometry was initially
estimated using photometry taken from a theoretical isochrone that fits the
cluster CMD, where we assumed that the system was composed of a bright, bluer
third star and two fainter, redder stars with identical characteristics.  
However, realistic
solutions for component A and the eclipsing stars cover large ranges 
(0.9 mag in $V$ for WOCS 24009 A, and 1.5 mag for the eclipsing stars),
showing this to be a weak constraint.

\begin{deluxetable}{ccccc}[!h]
\tablewidth{0pt}
\tablecaption{Secondary Eclipse Depths for WOCS 40007\label{a259depths}}
\tablehead{
\colhead{Filter} & \colhead{$\Delta m$} & \colhead{$N_{ecl}$} & \colhead{$N_{in}$} & \colhead{$N_{out}$} 
}
\startdata
$B$ & $0.3850\pm0.0031$ & 2 & 4,3 & 5,12 \\
$V$ & $0.4013\pm0.0042$ & 2 & 12,2 & 38,5\\
$R_C$ & $0.4215\pm0.0030$ & 3 & 7,1,4 & 19,9,5\\
$I_C$ & $0.4426\pm0.0071$ & 1 & 8 & 26
\enddata
\tablecomments{$N_{ecl}$ is number of observed eclipses, $N_{in}$ is the
  number of observations made during totality, and $N_{out}$ is the number of
  out-of-eclipse observations.}
\end{deluxetable} 

We were able to improve the decomposition with information from the broadening
functions (\S \ref{bf}) to get a fairly direct measurement of the relative
brightness of the three stars in a limited wavelength range within the $V$
passband. This would be an important constraint if the stars only underwent
partial eclipses. Closer examination of the light curves indicates that the
stars must at least come close to totally eclipsing because the eclipse depths
($\sim0.2$ mag) are consistent with the amount of light contributed by the
eclipsing stars in the broadening functions. Indeed, for the best resolved
eclipses, there is a brief period of totality, as could occur for stars
that differ in radius by less than 4\%. The light curve fits also indicate
that the inclination of the eclipsing binary is within a degree of being
edge-on. This news means that the system photometry can be precisely
decomposed: the light from component C in the eclipsing binary is directly
measurable from the secondary eclipse depth, the similarity of component B
along with light curve fits can then constrain its contribution, and the
bright component A is the remainder.

\begin{figure*}[!ht]
\centering
\includegraphics [width=0.7\textwidth]{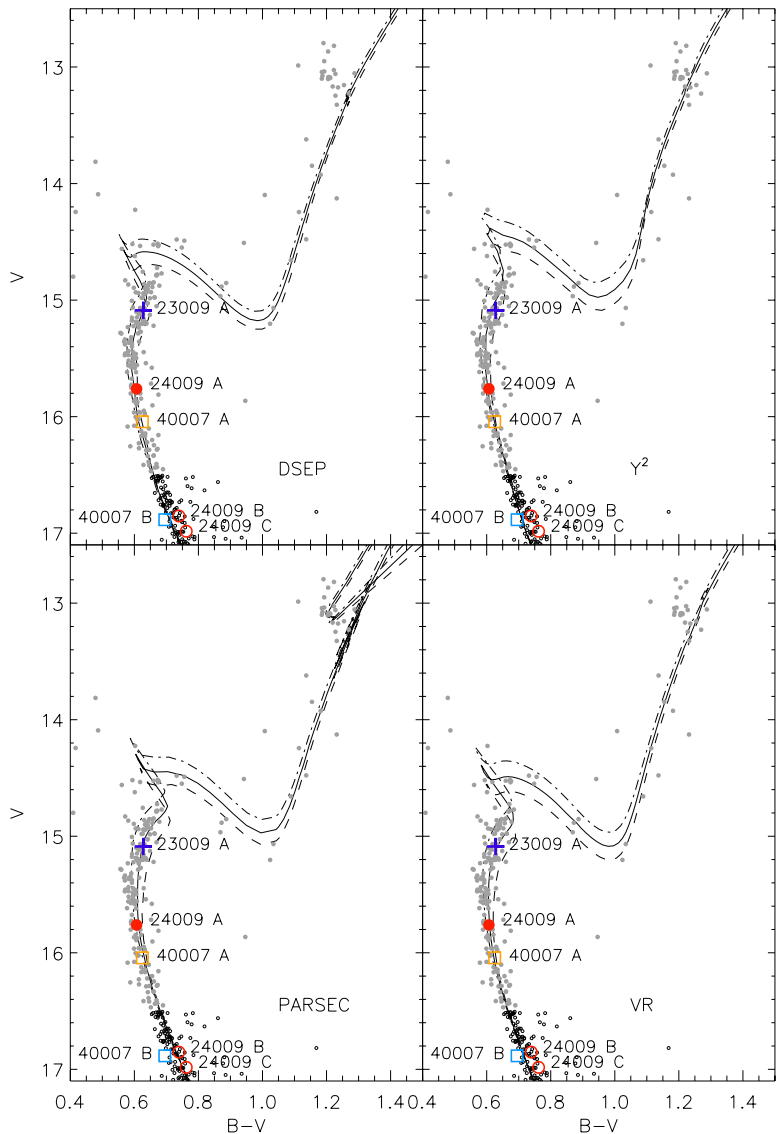}
\vspace{-15pt}
\caption{$BV$ CMD \citep{kalirai01} for NGC 6819 with removal of
  non-members having $V \leq 16.5$. Differential reddening corrections
  from \citet{platais13} are applied to each star. Isochrones are
  plotted with $Z \approx 0.015$ and ages 2.3, 2.5, and 2.7 Gyr for
  PARSEC and VR, while DSEP show 2.5, 2.7, 2.9 Gyr and Y$^2$
use 2.2, 2.4, and 2.6 Gyr; all use $E(B-V)=0.16$ and $(m-M)_V=12.40$.
  The decomposed photometry for WOCS 24009 is shown as the set of red
  points with the bright third star (filled circle) and the dimmer
  eclipsing stars (empty circles).  WOCS 23009 A (purple $+$) along with WOCS
  40007 A (orange $\square$) and B (blue $\square$) stars are indicated.
  \label{decomp_cmd}}
\end{figure*}

Because the period of totality is so brief and the contribution of
component A is so large, it is not as feasible to use a direct
measurement of the secondary eclipse depth in the same way as we did
for WOCS 40007 above. We therefore derived light ratios from our light
curve analysis and used them to decompose the system's
photometry. Although we did use the broadening function light
constraints in our models, the final light ratios are driven by the
existence of the total eclipse and the near identical characteristics
of the eclipsing stars.\footnote{For completeness, we note that the
  derived value of $l_3 = L_A / (L_A + L_B + L_C) = 0.614$. This
  underscores that the broadening function values from section
  \ref{bf} were systematically in error as a result of the significant
  temperature difference between component A and the eclipsing
  stars. However, the light curve fit overrode the constraint
  value we used.}  The decomposed photometry of WOCS 24009 is given in
Table \ref{summary}. The color of the bright A component is consistent
with the spectroscopic temperature for our preferred reddening, and
all three stars fall within the band of the main sequence in the
CMD. The decomposed photometry of WOCS 24009 can be found in Figure
\ref{decomp_cmd} along with the other measured cluster members.

Although we can find isochrones that fit the masses and photometry of
the eclipsing stars of each binary (WOCS 40007 and WOCS 24009)
separately to within $1\sigma$, there are no solutions that
simultaneously fit all four stars that precisely. A distance modulus
of $(m-M)_V \approx 12.45$ threads through the error ellipses for all
of the stars at a little more than $1\sigma$ away, but we must
consider the possibility of systematic errors. The relative
brightnesses of the stars in each binary are consistent with their
masses for models with a wide range of ages and compositions.
However, the components of WOCS 40007 imply a smaller distance modulus
than WOCS 24009 by about 0.1 mag. Differential reddening is a natural
place to seek an explanation for this discrepancy because the
\citet{platais13} map appears to be over-correcting WOCS 40007: both
components appear to be bluer than the corrected main sequence, and
WOCS 40007 B is significantly brighter than WOCS 24009 C in spite of a
smaller measured mass. However, to bring the WOCS 40007 stars back to
the cluster main sequence, the reddening correction would need to be
well outside of the distribution of corrections for nearby stars
(within about an arcmin).  Proper-motion and radial-velocity
information argues against the idea that WOCS 40007 is a foreground
nonmember system. Our current best explanation is that the WOCS 40007
star characteristics have been affected by star-star interactions in
the case of component B, and by imperfectly determined
corrections for the faint third star star in the case of component A.

\subsubsection{WOCS 23009}

The simplest case is the long period binary WOCS 23009. The small
amplitude of the total secondary eclipse in the {\it Kepler}
photometry \citep[0.002 mag;][]{sandquist13} indicates that we can treat
the system photometry as essentially identical to that of the brightest eclipsing
star --- any corrections would be less than the uncertainty in the
system magnitude.  In all of the measured DEBs in NGC 6819, WOCS 23009 A
is potentially the best indicator of age for the cluster, as it is the
most evolved star with a measured radius. 

\begin{figure*}[!ht]
\centering
\includegraphics [width=0.7\textwidth]{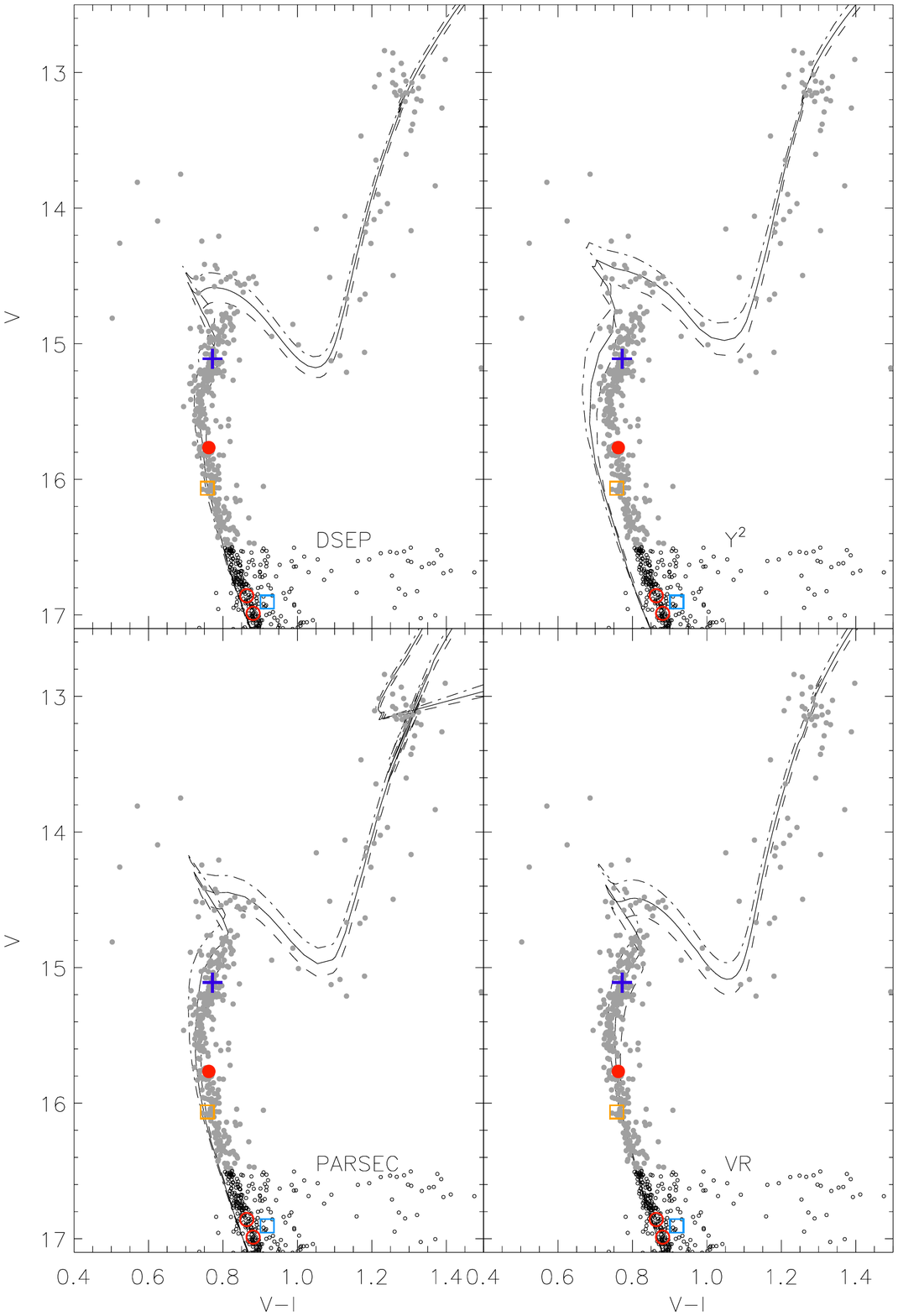}
\vspace{-15pt}
\caption{$VI_C$ CMD for NGC 6819 \citep{yang} with isochrones employing
  $E(V-I)=0.20$ and $(m-M)_V=12.40$. Otherwise the isochrone ages and
  symbols are the same as Fig. \ref{decomp_cmd}.
\label{yangcmd}}
\end{figure*}

We have updated our estimate of the mass of WOCS 23009 A using the
photometry and masses of our eclipsing binary star measurements.
\citet{sandquist13} determined the mass of WOCS 23009 A to be
$1.468\pm0.030 M_\sun$ using an isochrone-based extrapolation from the
masses and photometry of the components of the binary WOCS 40007. It
is almost certainly more massive than WOCS 24009 A, but it is part of
a single-lined binary and so its mass could only be estimated from its
CMD position relative to other cluster stars of known mass.  Our
preferred fit uses the two WOCS 24009 stars because of larger concerns
about the photometry of the WOCS 40007 stars.  Fits for a choice of
isochrone (from a particular research group with specific composition
and age) give uncertainties of about $\pm0.010 M_\sun$ for the mass of
WOCS 23009 A, with an additional $0.004 M_\sun$ coming from
uncertainty in the $V$ magnitude of WOCS 23009. However, there are
larger systematic differences. Using $Z=0.015$, Dartmouth (DSEP;
\citealt{dotter08}) and Victoria-Regina (VR; \citealt{vandenberg06})
isochrones produce similar masses ($1.472-1.476 M_\sun$), while the
PARSEC \citep{bressan12} models give a lower mass ($1.458
M_\sun$). The lower mass for the PARSEC isochrones may be due to a
larger assumed convective core overshooting (approximately 0.25
pressure scale heights $H_P$ versus $0.2 H_P$ for other models) or a
higher assumed helium abundance ($\Delta Y \approx 0.007$) for a given
metal content.  When varying the metal content along with the helium
content (assuming, like most stellar modelers, that there is a
constant $\Delta Y / \Delta Z$), there are smaller systematic changes
because of the compensating effects of $Y$ and $Z$: reducing the metal
content from $Z=0.015$ to $Z=0.012$ reduces the mass of WOCS 23009 A
by about $0.004 M_\sun$.  Based on these considerations, we estimate the systematic
uncertainty due to model differences to be $\pm0.008 M_\sun$.


Our new estimate of the mass of WOCS 23009 A is $1.464\pm0.011\pm0.012
M_\sun$, where the two uncertainties are estimates of the random and
systematic effects, and we have taken the central value to be between
the PARSEC and DSEP/VR values using $Z=0.012$ and
0.015. This is consistent with our original estimate ($1.468\pm0.03
M_\sun$, \citealt{sandquist13}). The main changes in the analysis are
the revision in the metal content of the models we use and the
different sample of stars used for the mass estimation. If we were to
use the stars of WOCS 40007 instead, we would get a mass for WOCS
23009 A that is about $0.02 M_\sun$ lower. We hope that in the future we
can reconcile the characteristics of WOCS 40007 with those of the
other eclipsing systems, but right now we believe there are still
unidentified systematic issues present.

Because the radius of the star is mildly correlated with the mass
estimate used \citep{sandquist13}, we have re-run fits to the radial
velocities and eclipse light curves of WOCS 23009. The updated values
are shown in Table \ref{summary}.

\begin{figure*}[!ht]
\centering
\includegraphics [width=0.70\textwidth]{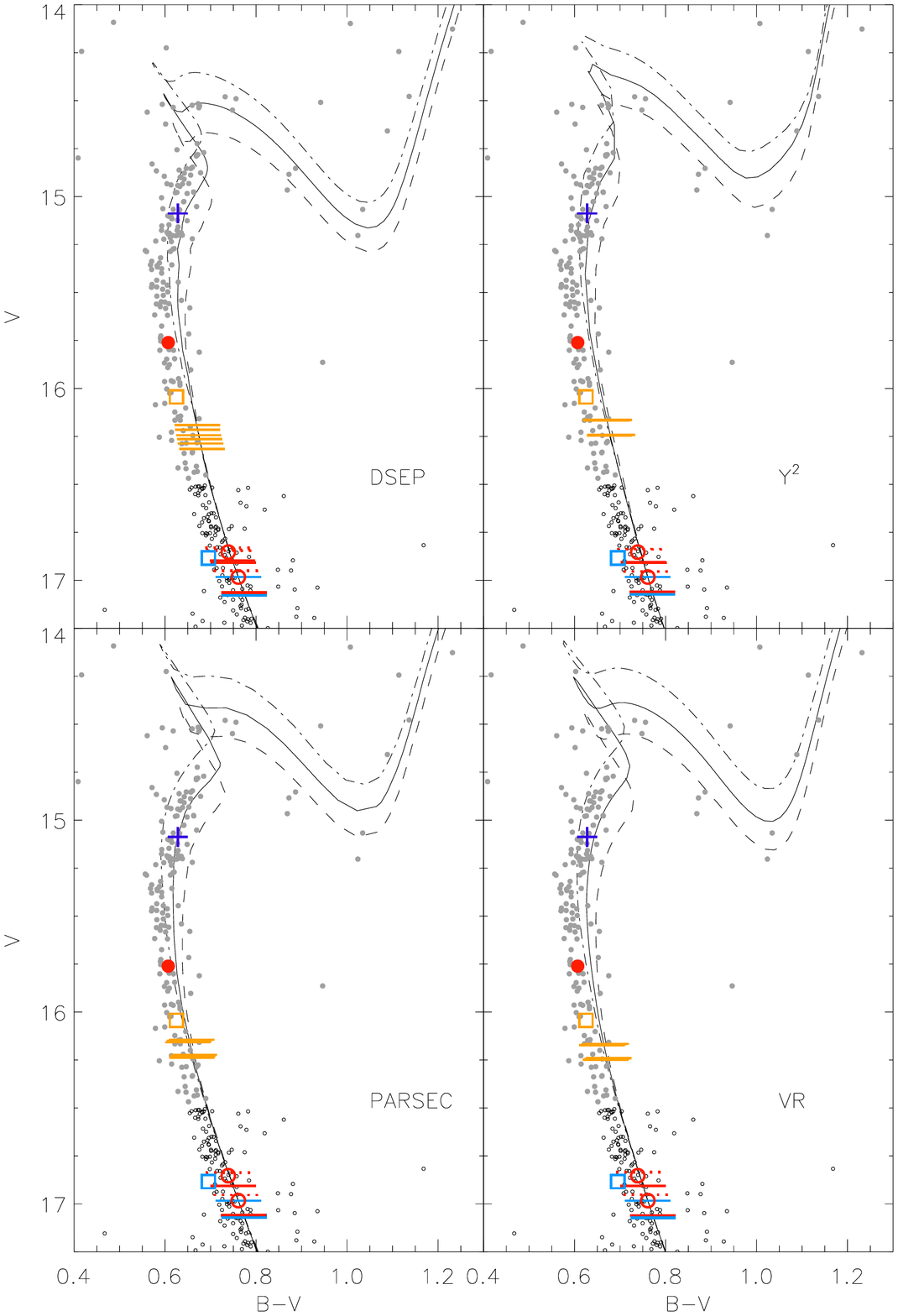}
\vspace{-15pt}
\caption{$BV$ CMD for NGC 6819 with $Z \approx 0.015$ isochrones
  shifted to match the photometry of WOCS 24009 C at its measured
  mass, as discussed in \S \ref{age}.  The symbols for the plotted cluster 
  stars are the same as Fig. \ref{decomp_cmd}. Horizontal lines delimit the
  range for the 1$\sigma$ uncertainty in the mass for components. Those for WOCS
  24009 are red dotted lines for component B and solid for component C. 
  Plotted ages are 2.1, 2.3, and 2.5 Gyr for DSEP and
  PARSEC; 2.0, 2.2, 2.4 for VR and Y$^2$.
\label{force}}
\end{figure*}

\subsection{Distance Modulus}

When radius and temperature can be determined for eclipsing binary
stars, the distance can be derived by comparing the predicted absolute
magnitude to photometric observations. We used the photometric
temperature estimates from \S \ref{temps} along with the binary star
radii to derive luminosities, applied bolometric corrections from
\citet{vandc} to compute absolute $V$ magnitudes, and used $V$
magnitudes from \citet{yang} photometry to derive the distance
modulus. For four out of the five eclipsing binary stars we discuss here (WOCS 23009 A,
WOCS 24009 B and C, and WOCS 40007 A) we get very consistent distance
moduli. For the remaining star (WOCS 40007 B), the difference is
likely due to problems determining the photometric temperature:
$(B-V)$ and $(V-I)$ colors for the star puts it on the blue and red
sides of the main sequence, respectively.  If we adopt a temperature
consistent with those of WOCS 24009 B and C (very close to WOCS 40007
B in the CMD), the distance modulus comes into agreement with those of
the other stars.

After propagating uncertainties due to the mean cluster reddening,
differential reddening, and the decomposition of the photometry of the
binary components, we find typical uncertainties of around 80-100 K in
the photometric temperatures. The uncertainty in the temperature
dominates the error budget for the distance modulus, which we find to
be about 0.08 mag for each star. The individual distance values are
given in Table \ref{summary}. The weighted average is
$(m-M)_V=12.38\pm0.04$, where the quoted uncertainty is the error in
the mean. This is consistent with the recent value ($12.40\pm0.12$)
derived by \citet{att} in a comparison of the CMD with Yonsei-Yale (Y$^2$; \citealt{yy})
isochrones.  Our distance modulus determination is less
explicitly dependent on the choice of metallicity, although it does
enter through the photometric temperature determination. However, a
0.1 dex shift in [Fe/H] only results in a 15 K change in temperature
and about a 0.01 mag shift in the distance modulus.

\begin{figure*}[!ht]
\centerline{\includegraphics [width=0.7\textwidth]{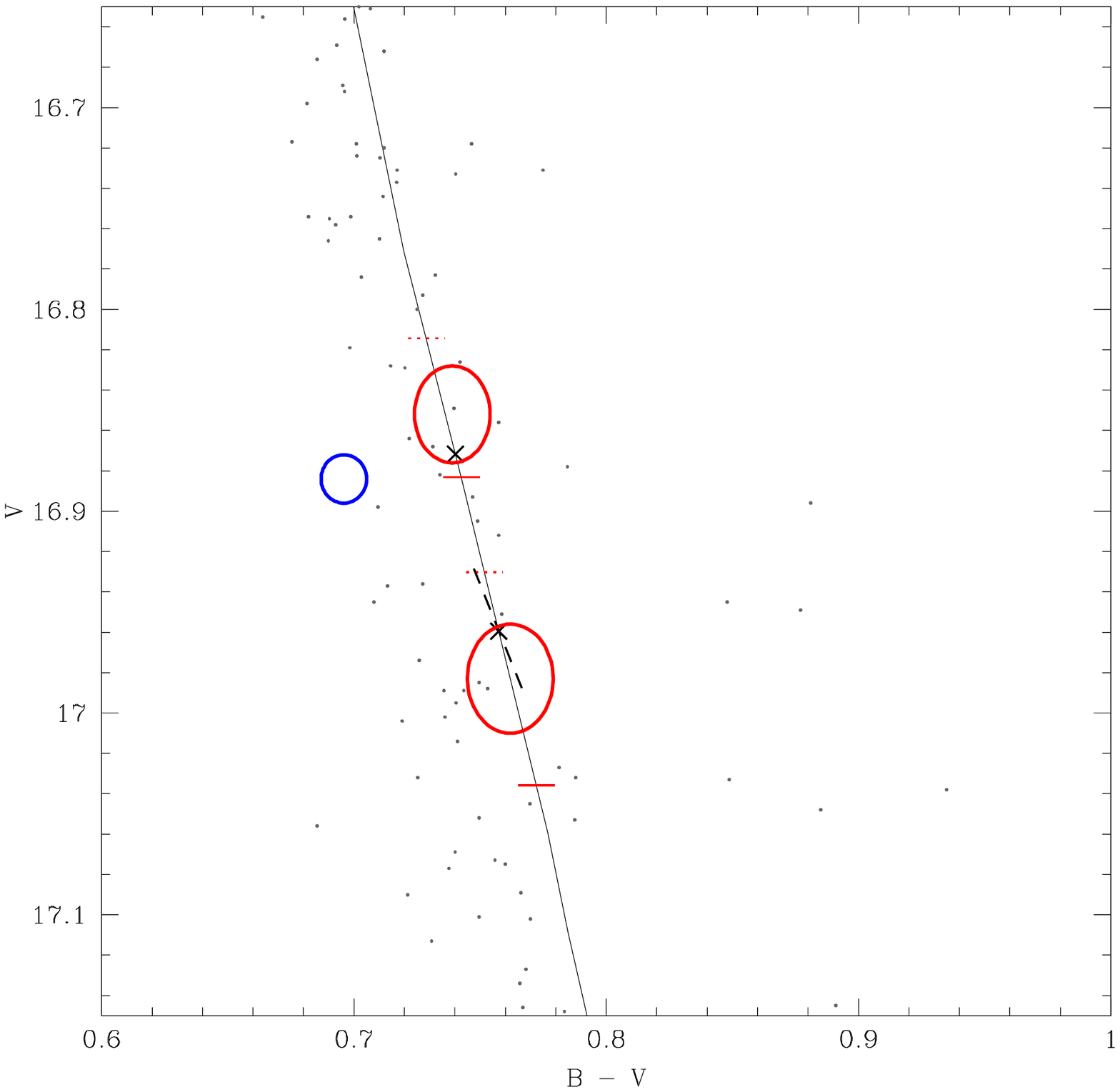}}
\vspace{-80pt}
\caption{$BV$ CMD for NGC 6819 zoomed on the fainter components of
  WOCS 24009 (red) and WOCS 40007 (blue), shown with a $1\sigma$ error
  ellipses for their photometry. A 2.3 Gyr $Z \approx 0.015$ PARSEC
  isochrone is plotted with the predicted positions of stars of masses
  equal to WOCS 24009 B and C ($\times$) and the $1\sigma$ mass
  uncertainties (horizontal lines).  The isochrone is shifted to
  optimize the matches to the photometry of WOCS 24009 B and C at
  their measured masses. The effects of a $\pm0.01$ mag uncertainty in
  the differential reddening color correction is shown with the black
  dashed line.
\label{faintzoom}}
\end{figure*}

\subsection{Age Constraints}\label{age}

When fitting isochrones to cluster CMDs, age
information comes from the photometry of stars around the main
sequence turnoff. As a baseline comparison (Fig. \ref{decomp_cmd}), we first employed the conventional method
of comparing isochrones to cluster photometry using a reddening and distance modulus,
in this case determined by \cite{att}. We used the
photometry of \citet{kalirai01} due to its high signal-to-noise ratio
(and small scatter) for main sequence stars near the turnoff.  To
minimize the scatter, we also applied the differential
reddening correction map used by \citet{platais13}. We further cleaned
the CMD by restricting the sample to single-star cluster members
brighter than the $V \sim$ 16.5 magnitude limit from the WOCS RVs
survey \citep{milliman} and to stars with proper-motion membership
probabilities greater than 50\% from either \citet{platais13} or
\citet{sanders}. The decomposed photometry of WOCS
24009 is plotted along with the photometry of other observed DEB
stars: WOCS 23009 A \citep{sandquist13} and both eclipsing components
of WOCS 40007 \citep{jeffries13}. The tabulated photometry of the
cluster DEB stars can be found in Table \ref{summary}.  To facilitate
clean comparisons between isochrone sets as discussed at the beginning
of \S \ref{results}, we use a common value of $Z \approx 0.015$ for
all of the isochrone sets: DSEP, PARSEC, VR, and
Y$^2$.  Figure \ref{yangcmd} uses $V$ and
$I_{c}$ photometry from \cite{yang} and also applies differential
reddening corrections to the CMD.  The
isochrones do a decent job of fitting the $VI_{c}$
photometry along the main sequence, turnoff, and subgiant branches
except for Y$^2$ models, which appear to be too blue for the cluster stars
as well as PARSEC models to a lesser extent.
Due to distance and reddening uncertainty for the cluster, forcing
isochrones to fit well-measured stars at their observed $V$ magnitude,
color, {\it and} mass is a more robust way of comparing isochrones to
the CMD.  We chose to align the models with WOCS 24009 C in
Fig. \ref{force} because of its precise decomposed photometry, and by
doing so there is no need to assume a reddening and distance modulus.
This type of differential comparison, where we shift all isochrones to match 
mostly unevolved main sequence stars, uses the isochrone shape at the 
turnoff and brightness as the age indicator. 
In the comparison here, the isochrones do a satisfactory, if
imperfect, job of matching the characteristics of the fainter
eclipsing stars.  For example, isochrones predict that a star with the
measured mass of WOCS 40007 A should be fainter relative to the three
stars with mass in the range $1.005 < M / M_\sun < 1.010$ --- compare
the inferred position of the star with the range predicted based on
its measured mass (delimited by horizontal lines). However, there are
lingering uncertainties in the photometric decomposition of WOCS 40007
A due to the faint third star (described in \S \ref{bf}) that may
alleviate this issue. Future refinements of the mass of WOCS 24009 A
should help identify whether this is a serious issue with the models.

\begin{figure*}[!ht]
\centering
\includegraphics [width=0.70\textwidth]{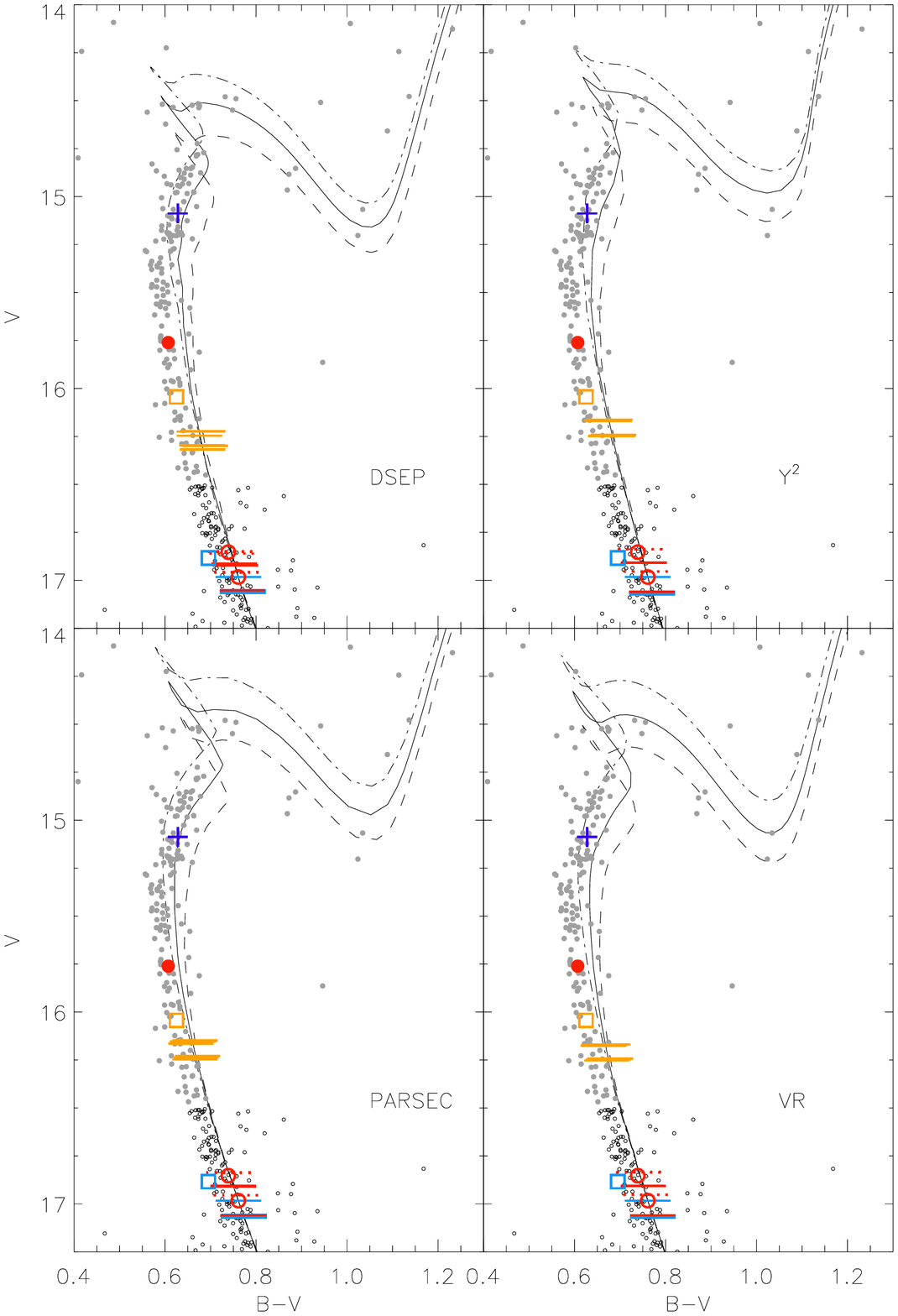}
\vspace{-15pt}
\caption{$BV$ CMD \citep{kalirai01} for NGC 6819 with $Z \approx
  0.012$.  All isochrones are shifted in the same way as in
  Fig. \ref{force}, and symbols are the same. Plotted isochrone
  ages are 2.0, 2.2, and 2.4 Gyr for DSEP, PARSEC, VR, and Y$^2$.
  \label{force01}}
\end{figure*}

The most age sensitive features of the CMD at present are the
photometric positions of WOCS 23009 A and the stars on the turnoff and
subgiant branch. WOCS 23009 A is valuable because its companion
is rather faint, so that we can be assured that its measurements
accurately represent the photometry of a single star. Unfortunately, we
cannot directly measure its mass. In addition, there are uncertainties
in the fit due to uncertainty in the mass and photometric decomposition of WOCS 24009 C, 
and in the relative
differential reddening corrections for WOCS 23009 A and 24009 C. All
of these affect where the isochrones should be
pinned at the faint end. The mass uncertainty is the dominant
uncertainty (see Figure \ref{faintzoom}), leading to an uncertainty of
about $\pm0.08$ mag in the isochrone placement. The fit to the
combination of WOCS 24009 B and C can be improved by a brightward
shift of 0.02 to the isochrones --- both components can be matched to
within the uncertainties on the masses and decomposed photometry, as
shown in Figure \ref{faintzoom}. Regardless of that choice, these
uncertainties together lead to an uncertainty of about 0.1 Gyr in the
age implied by WOCS 23009 A.  Based on the comparison with $Z \approx
0.015$ models, the
implied ages are 2.12 Gyr for VR, 2.15 Gyr for Y$^2$, 2.21 Gyr for
DSEP, and 2.31 Gyr for PARSEC. Based on the scatter in these values,
we estimate a systematic uncertainty of about 0.09 Gyr due to
differences in the physics encoded in the models.

\begin{figure*}[!ht]
\centering
\includegraphics [width=1\textwidth]{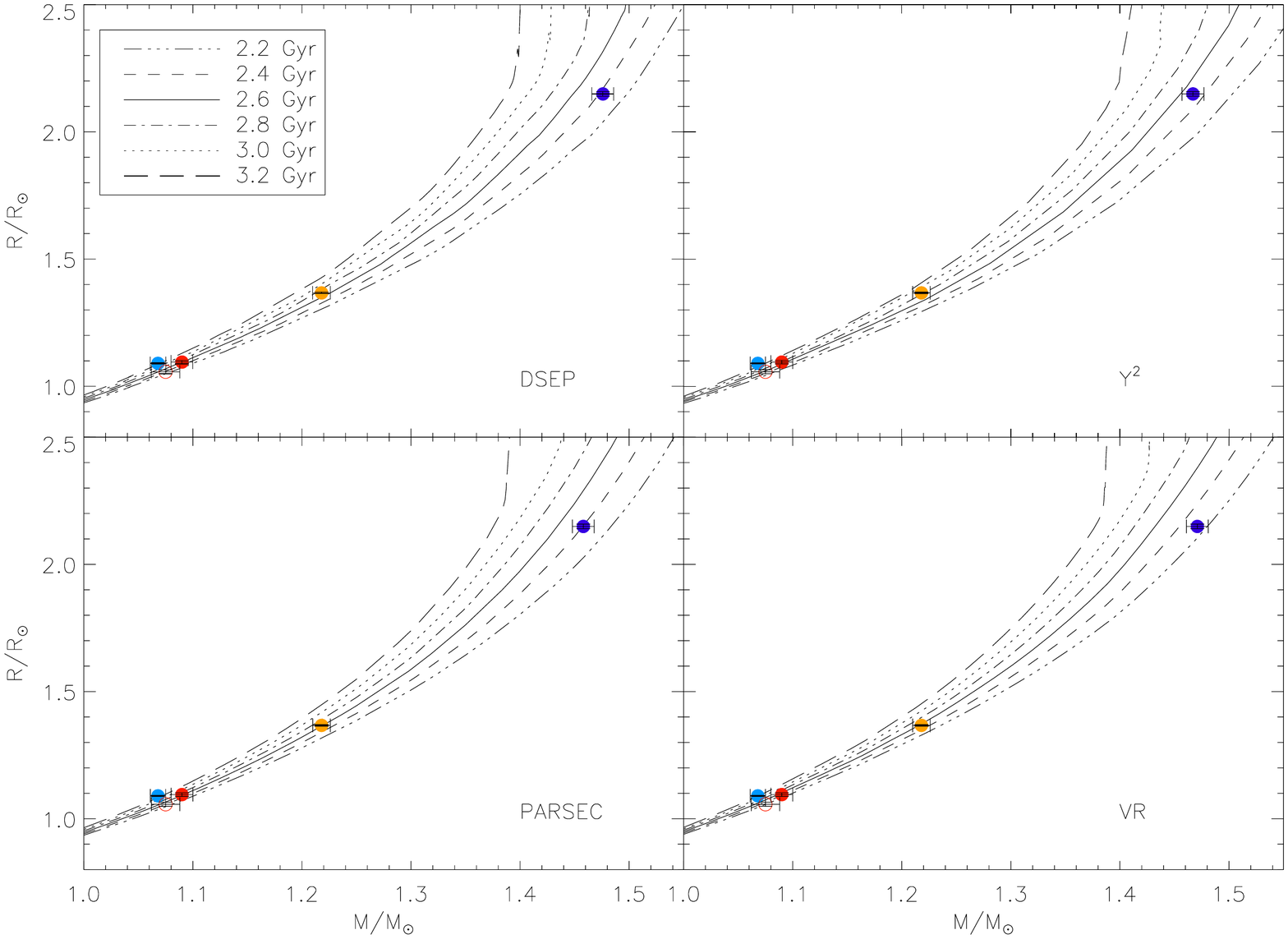}
\vspace{-25pt}
\caption{Mass-radius ($M-R$) plot for WOCS 24009 B and C (red filled and empty circles),
  WOCS 40007 A (orange) and B (blue), and WOCS 23009 A (purple) as the most massive component. 
  The plotted isochrones employ $Z = 0.015$ and [$\alpha$/Fe] = 0.0. The
  masses used for WOCS 23009 A are derived from extrapolations using
  the plotted isochrone set.
   \label{mr09}}
\end{figure*}

The turnoff and subgiant stars are more numerous and contain age
information, but we have imperfect knowledge of whether they might be
in undetected multiple star systems. At the turnoff (the bluest point
on the isochone) and fainter, the blue side of the observed 
distribution of main sequence stars can usually be expected to be a good
representation of the single star main sequence because unresolved
binaries would be displaced brighter and redder. Brighter than the
turnoff, however, unresolved binaries can fall either redder or bluer
than the single star sequence. On the subgiant branch, binarity can
affect the color in either direction, but the lower envelope of the
subgiant stars can trace out the single star sequence. Using
proper-motion and radial-velocity information from \citet{milliman},
we plotted the most likely single star subgiant cluster members on all
of our CMDs: WOCS 7008, 8004, 9009, 11005, 11006, 18002, 20016, 22004,
22013, and 23006.  Satisfactory fits are possible with each of the
isochrone sets, but the color of the turnoff and the brightness of the
faint envelope of the subgiant branch appear to be poorly matched.
Part of this may be uncertainties in the measured positions of the
faint eclipsing stars we use to pin the isochrones, but the shapes of
the isochrones on the subgiant branch and at the top of the main
sequence can be importantly affected by the modeling of convective
core overshooting. For example, the DSEP models appear to do the best
job of simultaneously matching the properties of WOCS 23009 A and the
fainter subgiant branch stars. Without additional work on the
subgiants and their modeling, it is too early to weight them in the
determination of the age presented here.

Uncertainties in metallicity also produce a potential source of
systematic error though. We examined isochrone fits using a decreased
metal abundance $Z=0.012$ in Figure \ref{force01}. When using a lower
metal abundance, the isochrones used to best fit the cluster are
younger by about 0.1 Gyr compared to isochrones using $Z=0.015$. The
main sequence shape seems to be somewhat more poorly fit for all four
isochrone sets. From all of the considerations of the CMD, we quote an
age of $2.21\pm0.10\pm0.20$ Gyr, where the quoted uncertainties are
from the combined fitting uncertainties and the combined systematic
uncertainties.  The shifts needed to match the photometry of WOCS
24009 C at its measured mass imply particular values of reddening and
distance modulus, and for completeness, we give them here. For the
four sets of isochrones we have discussed, we find reddenings $0.23 <
E(B-V) < 0.26$ and distance moduli $12.56 < (m-M)_V < 12.67$ for
$Z=0.012$, and $0.19 < E(B-V) < 0.22$ and $12.46 < (m-M)_V < 12.57$
for $Z=0.015$. These values are affected by differences in input
physics and color-temperature relations used for the different
isochrone sets, and we do not consider them to be measurements of
these values.

\begin{figure*}[!ht]
\centering
\includegraphics [width=1\textwidth]{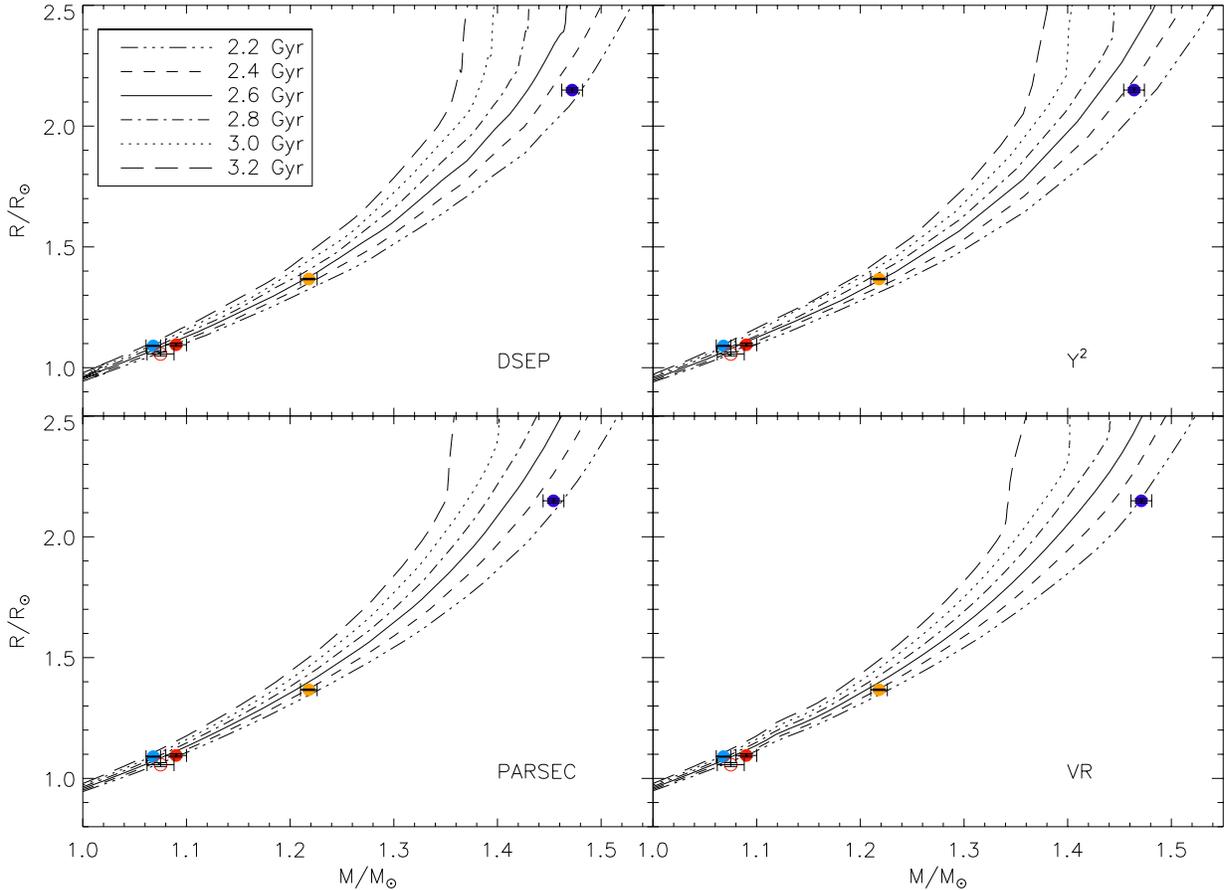}
\vspace{-25pt}
\caption{$M-R$ plot for isochrones with a lower metal abundance $Z = 0.012$.
Star symbols are the same as in Fig. \ref{mr09}.
\label{mr01}}
\end{figure*}

The most direct method of comparing eclipsing binary observations to
theory is through the mass-radius ($M-R$) plane. The radii of evolved
stars are good age indicators because the physical measurements in an
eclipsing binary system can reach a high degree of precision while
avoiding sources of systematic error from sources such as distance and
reddening.  Statistically speaking, it is best to use all
well-measured stars for determining the cluster age, and with the
analysis of WOCS 24009, we add two more stars. (Unfortunately, WOCS
24009 A does not undergo eclipses, and so its radius cannot be
measured directly and it cannot be used in this analysis.)  The masses
and radii for WOCS 24009 and the other cluster stars used in determining age are
listed in Table \ref{summary}. Note that the values
for WOCS 40007 A and B have been updated for this paper after improved
corrections for the effects of the faint third star on RVs and
eclipse timings, and the incorporation of {\it Kepler}
photometry. Figure \ref{mr09} shows the $M-R$ diagram comparison with
isochrones assuming the higher metal abundance $Z = 0.015$.  For all
model sets using the higher metal content, the eclipsing stars of WOCS
40007 lie slightly above the isochrone that best matches the bright
star WOCS 23009 A and the eclipsing stars 24009 B and C.  Based on the
\cite{att} suggestion of a lower metallicity, we examined a metal
abundance $Z = 0.012$ (decreased by about 0.1 dex and consistent with
plausible errors in the metallicity scale) as seen in Figure \ref{mr01}.
With a lower metallicity the eclipsing stars fall on younger models,
similar to what is seen in the CMDs with the same $Z$ content. The
model radii decrease and produce a small improvement to the
consistency of the ages implied by the different stars and a small
($\sim$0.2 Gyr) reduction to the age.

The most age-sensitive eclipsing star currently known in the cluster
is WOCS 23009 A, and the age uncertainty derived from this star alone
is mostly responsible for the error in the weighted mean. It should be
remembered that its mass comes from an isochrone-based extrapolation
from the properties of WOCS 24009 stars as mentioned in \S
\ref{debphot}, but that the radius is not terribly sensitive to this
estimate (see \citealt{sandquist13} for a discussion) and this is why
this star can provide age discrimination.  With the improved radial
velocity measurements presented here, the mass uncertainties for the
stars in WOCS 40007 and WOCS 24009 have been reduced compared to what
was presented in \citet{jeffries13}. The eclipsing stars for those two
systems appear to imply slightly different ages, although this is not
at a high level of significance. For example, the three faintest stars
(WOCS 40007 B and WOCS 24009 B and C) have masses that are consistent
with each other at a little under a $2\sigma$ level. As discussed in
\S \ref{inflate}, WOCS 40007 B may have been inflated in radius due to
tidal interaction, and its V-band photometry may also have been
affected, making its use in the CMD age determination
questionable.  Depending on the isochrones used, the remaining three
stars are roughly consistent with the age derived from WOCS 23009 A at
about the $1\sigma$ level.

The error in the weighted average is about 0.05 Gyr, but the
systematic sources of error are more substantial. A major systematic
contribution is uncertainty in the metal content of the stars. We find
differences of $0.13 - 0.20$ Gyr between models having $Z = 0.012$ and
$Z = 0.015$ from the same research group, with the lower metallicity
producing lower ages. As a result, we estimate the systematic
metallicity uncertainty to contribute an uncertainty of $\pm0.10$ Gyr
to the age. Comparing results from different research groups for the
same $Z$, there is a $0.2 - 0.25$ Gyr range, with VR
models giving the lowest ages (2.22 Gyr for $Z = 0.0125$), Y$^2$
models giving the highest (2.40 Gyr for $Z = 0.012$), and DSEP and
PARSEC models returning values nearly midway between these
values. Based on this, we estimate a systematic uncertainty of about
$\pm0.12$ Gyr on the age due to differences in physics and chemical
mixes used in the different models.
 
Because we do not have strong reason to prefer one metal content $Z$
or one set of models, we take our best $M-R$ estimate of the age to be
in the middle of the ranges and quote an age of $2.38\pm0.05\pm0.22$
Gyr, where the two uncertainties are statistical and systematic,
respectively. Although we can expect an improvement of the statistical
uncertainty in the age with additional measurement of the eclipsing
stars, the best way of improving the age estimate in the cluster will
be to reduce the metallicity uncertainty (through a combination of
improved spectroscopic measurements of cluster star abundances
relative to the Sun and a resolution of the question of the solar
value $Z_\sun$) and the model isochrone uncertainties.

\section{Conclusions}

We have presented extensive photometric and spectroscopic observations
of the detached eclipsing binary WOCS 24009 in open cluster NGC 6819
as part of a larger effort to obtain the age of the cluster using all
of the useful binaries (including WOCS 40007 and WOCS 23009). Our modeling indicates that we have obtained
mass and radius measurements with uncertainties $\sim$ 1.0\% and
0.7\% for the eclipsing components of WOCS 24009. We have started to characterize the
turnoff-mass WOCS 24009 A, although the uncertainty in its mass is
about 4.6\%.  This uncertainty can be reduced with a combination of
radial-velocity measurements of component A and eclipse timings
--- together these measurements will provide the highest precision
measurement of the third star's orbit and the mass ratio $q_A = M_A / (M_B
+ M_C)$.  Because the star's photometry can be disentangled from the
eclipsing stars, a more precise mass will provide a significantly
better constraint on the cluster age.

Systematic uncertainties in the theoretical stellar models and in the
metal content of the cluster stars are the main obstacles to a higher
precision age at present. We derive ages for NGC 6819 of
$2.21\pm0.10\pm0.20$ Gyr using the CMD and $2.38\pm0.05\pm0.22$ using
the $M-R$ plane, where the quotes are estimates of statistical and
known systematic uncertainties.
There is still significant disagreement between different indicators
of the cluster metallicity, and a better determination of this would
significantly reduce the systematic errors quote above. The possibility
of a lower cluster metallicity can be spectroscopically tested in the future 
using the binary stars with the assistance of constraints on the
surface gravities from the binary star analysis.  Our derived age is
in agreement with asteroseismic \citep{basu11} and white dwarf cooling
\citep{bedin} ages for the cluster.

Stellar model differences remain one of the biggest contributors to
the uncertainty in the age. To reduce this source of error, it is
helpful to review the systematic differences between model sets even
if we cannot be completely certain of the causes. In the $M-R$
comparisons made at the same metal content $Z$, VR
isochrones gave the youngest ages of the model sets we tested, and
Y$^2$ returned the largest ages. In CMD comparisons, VR,
Y$^2$, and DSEP models gave similar ages, with PARSEC models returning
the oldest age. The dependencies of these comparisons on different
model inputs (chemical composition mix, the treatment of core
convection and diffusion, and color-temperature transformations, among
others) should continue to help drive us toward more accurate models.

\acknowledgements We would like to thank the former Director of Mount Laguna
Observatory (P. Etzel) for generous allocations of observing time;
A. Talamantes, E. Bavarsad, and D. Baer for assistance with data taking at
MLO; and D. Short for his calculations testing the three-body orbital dynamics
of this system.

This paper includes data collected by the {\it Kepler} mission. Funding for
the {\it Kepler} mission is provided by the NASA Science Mission
directorate. We are grateful to the {\it Kepler} team for the opportunity to
work with such an extensive and precise dataset. We gratefully acknowledge
funding from the National Science Foundation under grant AST-0908536 to San
Diego State University and AST-0908082 to the University of Wisconsin-Madison,
and from the National Aeronautics and Space Administration (NASA) under grant
G00008957. 
K.B. acknowledges funding from the Carlsberg Foundation and the
Willum Foundation. Funding for the Stellar Astrophysics Centre is provided 
by The Danish National Research Foundation (Grant agreement no.: DNRF106). 
The research is supported by the ASTERISK project (ASTERoseismic 
Investigations with SONG and Kepler) funded by the European Research Council 
(Grant agreement no.: 267864).
A.M.G. is funded by a National Science Foundation Astronomy 
and Astrophysics Postdoctoral Fellowship under Award No.\ AST-1302765.

\bibliographystyle{plain}

\end{document}